\documentclass[a4paper,10pt]{article}

\usepackage{authblk}
\usepackage{graphicx}
\usepackage{makeidx}
\usepackage{bigstrut}
\usepackage{fancyref}
\usepackage{listings}
\usepackage[titletoc]{appendix}
\usepackage{titlesec}
\usepackage{indentfirst}
\usepackage{subfigure}
\usepackage{caption}
\usepackage{amssymb}
\renewcommand{\figurename}{Fig.}
\renewcommand{\tablename}{Table}
\newcommand*{\figref}[1]{\figurename~\ref{#1}}
\newcommand*{\tabref}[1]{\tablename~\ref{#1}}

\titleformat{\section}{\normalsize\bfseries}{\thesection}{1em}{}
\titleformat{\subsection}{\normalsize\itshape}{\thesubsection}{1em}{}
\titleformat{\subsubsection}{\normalsize\itshape}{\thesubsubsection}{1em}{}

\author[1]{A.~Bueno\thanks{a.bueno@ugr.es}, A.~Gasc\'on\thanks{agascon@ugr.es}}
\affil[1]{\footnotesize{Dpto. de  F\'isica Te\'orica y del Cosmos \&
    CAFPE, Universidad de Granada, E-18071, Granada, Spain}}
\title{CORSIKA Implementation of Heavy Quark Production and
  Propagation in Extensive Air Showers}

\makeindex

\begin{document}
\maketitle

\noindent \textbf{Abstract}\newline

Heavy quarks are commonly produced in current
accelerator experiments. Hence it is natural to think that they should be
likewise created in collisions with larger center of mass energies like
the ones involving ultra-high energy cosmic rays and atmospheric
nuclei. Despite this fact, a detailed treatment of heavy hadrons is missing in 
Monte Carlo generators of Extensive Air Showers (EAS). It is a must to
improve the description of how heavy
flavours appear and evolve in atmospheric showers. With this goal in mind, we study two different models for heavy quark
production in proton-air collisions. We also analyze a dedicated treatment of heavy hadrons interactions 
with atmospheric nuclei. This paper shows how those models have been
implemented as new options available in CORSIKA, one of the most used
EAS simulators. This new computational tool allows us to analyze the effects that the propagation of heavy
hadrons has in the EAS development.\newline

\noindent \textbf{Program summary}\newline

\noindent \textit{Title of program:} corsika-6990-Heavy\newline
\textit{Computer on which the program has been thoroughly tested:} Intel-Pentium based Personal Computers\newline
\textit{Operating system:} Linux\newline
\textit{Programming language used:} FORTRAN77 \newline
\textit{Memory required to execute:} 373 Mb\newline
\textit{Other procedures used:} NUCOGE [Linkai Ding, Evert Stenlund, Comput. Phys. Commu. 59 (1990) 313] \newline
\textit{Nature of physical problem:} Charmed and bottom hadron production and propagation inside Extensive Air Showers. \newline
\textit{Solution method:} Heavy quarks are produced according to two different production models. Propagation in the atmosphere
is handled using already present CORSIKA subroutines. New subroutines are written to simulate their interactions with
air nuclei.\newline
\textit{Restrictions on the problem:} Heavy quark production has only been implemented in the first interaction. \newline
\textit{Running time and output file size:} From 4.2 h and 120 Mb at $10^{19}$ eV 
 to 4.7 h and 170 Mb at $10^{19.75}$ eV, with thinning $10^{-6}\cdot$E(GeV).\newline

\newpage

\section{Introduction}
\label{sec:introduction}

Charm and bottom quarks are copiously produced at accelerators
(\cite{RHICHeavy,LHCbHeavy,Schieck,BabarHeavy}) and the 
physics involved in their production and hadronization processes is
reasonably understood \cite{PDG2012, Lourenco}. 
If they are produced in the energy range probed at colliders, we expect
them to be produced in the hadronic collisions taking place in
Extensive Air Showers (EAS) too, since the most
energetic cosmic rays reach energies of a few tens of EeV. At $\gtrsim$ 0.01 EeV heavy hadrons 
reach their critical energies and their decay probabilities decrease rapidly.
Decay lengths grow to considerable values: at 10$^{17.5}$~eV they are of the order $\sim 10$~km
for charmed and bottom hadrons. In addition, due to their larger masses, we expect heavy  
hadrons interactions with other hadrons to be more elastic on average, keeping a higher fraction of their 
energy after each interaction. Thus, above 0.1 EeV we expect 
the behavior of heavy hadrons to be very different from that at lower energies. 

Heavy quarks can in principle be produced at any stage of the
shower development, but it is only during the first interactions that
they can be produced with a significant
amount of energy (namely above their critical energies). Under those
circumstances, they can penetrate deep into the atmosphere giving rise
to additional contributions to the development of EAS.

Current Monte Carlo air shower simulators lack a full treatment of heavy hadrons. Charm production
is not always addressed, and charmed particle propagation is to a large extent neglected. 
As for bottom particles, they are neither produced nor propagated and they are not included 
in the list of particles considered for simulation.

In this paper we address the question of how to implement the physics
of heavy quarks inside the CORSIKA air shower simulator. In section \ref{sec:physics}
we summarize the production and propagation models
we use for the explicit treatment 
of charmed and bottom hadrons in EAS. 
The structure of the simulation chain is discussed in section \ref{sec:simchain}.
In section \ref{sec:effects} we analyze the effect of these modifications in the 
shower development. Details of the code implementation are presented
in the appendices. 
Throughout this work we use the following programs and software packages:
\begin{itemize}
 \item 	EAS are simulated using the CORSIKA version 6.990 \cite{corsika}, with:
 \begin{itemize}
  \item QGSJET01c to treat high energy interactions \cite{qgsjet01c}.
  \item FLUKA (version 2011.2.6) as the model for low energy interactions \cite{fluka1,fluka2}.
 \end{itemize}
 \item the number of nucleons participating in hadron-air collisions is simulated using NUCOGE \cite{NUCOGE}.

\end{itemize}

\section{Physics of heavy hadron production and propagation}
\label{sec:physics}

\subsection{Production models}
\label{sec:production}

The production of heavy quarks at accelerator energies is explained by Quantum Chromodynamics.
At cosmic rays collisions we are confronted with the problem of the energy scale, 
with collisions whose center of mass energies are of the order of 100 TeV. Accelerator data is several orders
of magnitude below and therefore extrapolation over large ranges of energy is mandatory.
However these extrapolations are prone to large uncertainties, as large as $\sim$40\% \cite{bCSuncertainties}.
To circumvent this difficulty, at the highest energies analysis based on effective theories are derived.
The solution to this problem is not unique, and different regimes allow for different approaches,
which in turn offer qualitatively correct results in certain limits. The Dual Parton Model \cite{DualPartonModel}, 
the Lund Fragmentation Model \cite{andersson} or the Color Glass Condensate model \cite{PartonSaturation} 
are some of the most well known effective theories. The latter is one of the most complete, supported by its analytic equivalence
with the gluon-gluon fusion mechanism of the parton model \cite{Equivalence}. 

However, not all features present in data are explained by this model.
Leading particle asymmetries have been reported from different fixed target experiments. They show a strong correlation between 
the quantum numbers of the projectile and those of the final state hadron, whereas according to the QCD factorization 
theorem heavy quarks hadronize independently of the initial state \cite{factorization}.
For example, in $\pi^-$($\bar{u}d$) interactions with hadrons or nuclei, the $D^-$ ($\bar{c}d$) carries on average a larger fraction 
of energy than the $D^+$ ($c\bar{d}$) \cite{BottomVogt,leadC}. Yet, the prediction stands 
that $c$ and $\bar{c}$ quarks should be produced with identical energy distributions. 

To explain this discrepancy theoretical models coincide in invoking a charm or bottom
component inside the nucleon. The Meson-Cloud model \cite{MesonCloud}, the Recombination Mechanism \cite{intrinsic_anjos}, 
or the Intrinsic Quark mechanism \cite{leadC,IntrinsicCB,IntrinsicB} are examples of these models, yet the nature 
and evolution of the heavy component differs between them. They produce similar results, but we will focus on the 
last model, which provides a simple mechanism for producing the flavor correlations present in data.


\subsubsection{Color Glass Condensate}
\label{subsec:CGC}

\begin{figure}[!t]
 \begin{center}
  \includegraphics[width=0.49\textwidth]{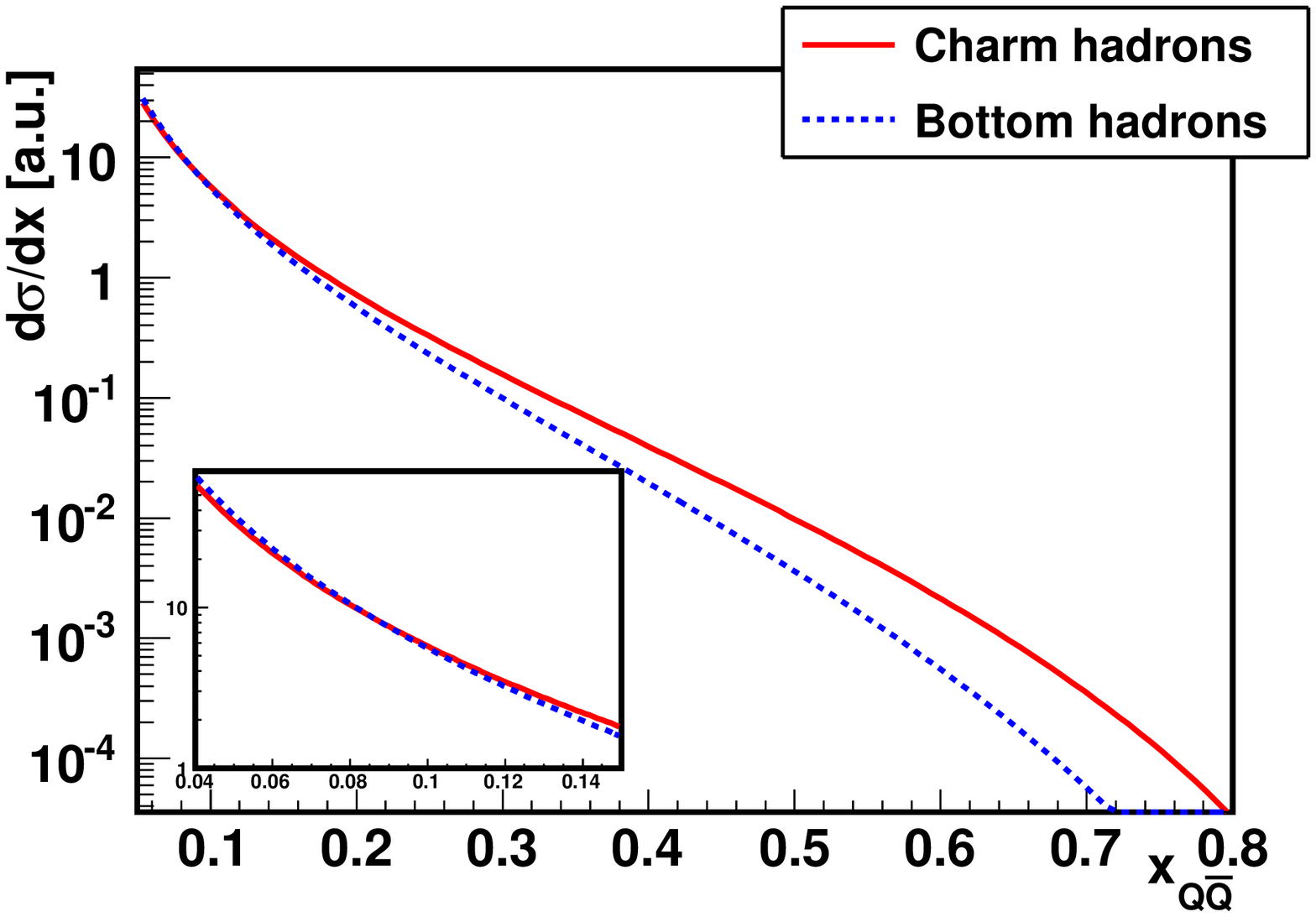}
  \hfill
  \includegraphics[width=0.45\textwidth]{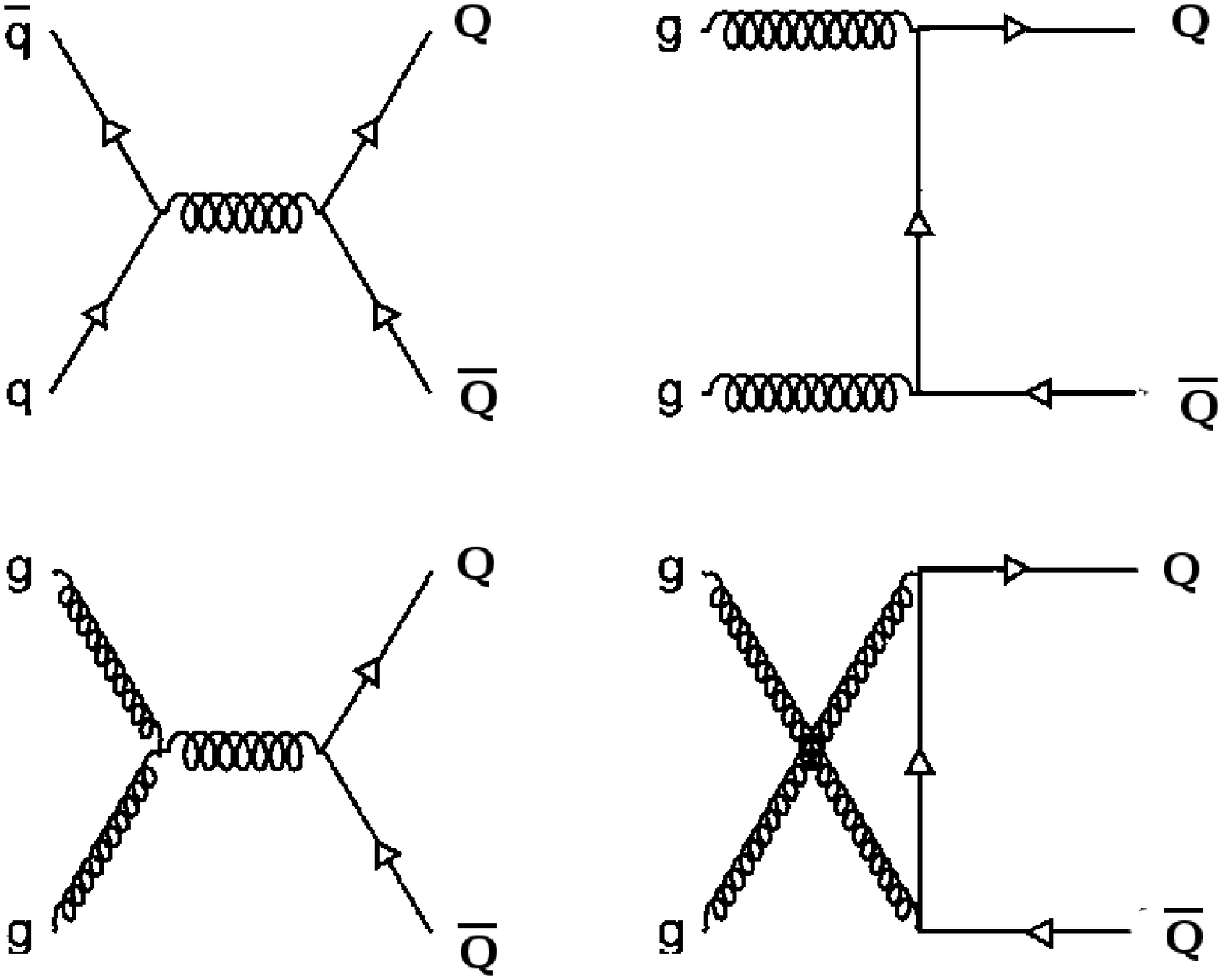}
 \end{center}
 \caption{Left: Differential fraction of primary energy carried by
   heavy hadrons produced in the Color Glass Condensate model. The
   inset zooms the region where the heavy quarks carry a small
   fraction of the initial energy.
 Right: Heavy flavor hadron production at leading order.}
 \label{fig:xfCGC}
\end{figure}

This is the first mechanism of heavy quark production we will briefly discuss and implement in CORSIKA (a
more technical discussion of the model can be found in~\cite{JHEP2007machado_erratum,Cazaroto}).
In this model, a heavy flavor quark-antiquark pair is created through the fluctuation of the probing gluon. 
Charmed and bottom hadrons are formed from hadronization of those heavy quarks with sea quarks, in a 
mechanism called Uncorrelated Fragmentation. Any heavy hadron has the same probability of
being formed from the heavy quarks produced. We assume that hadronization occurs without
energy loss, and thus the differential production probability for charmed (bottom) hadrons is identical
to that of charm (bottom) quarks. Those distributions can be seen in \figref{fig:xfCGC} (left), 
both scaled to the same integral.

\subsubsection{Intrinsic Quark production}
\label{subsec:IQ}

\begin{figure}[!t]
 \begin{center}
   \includegraphics[width=0.49\textwidth]{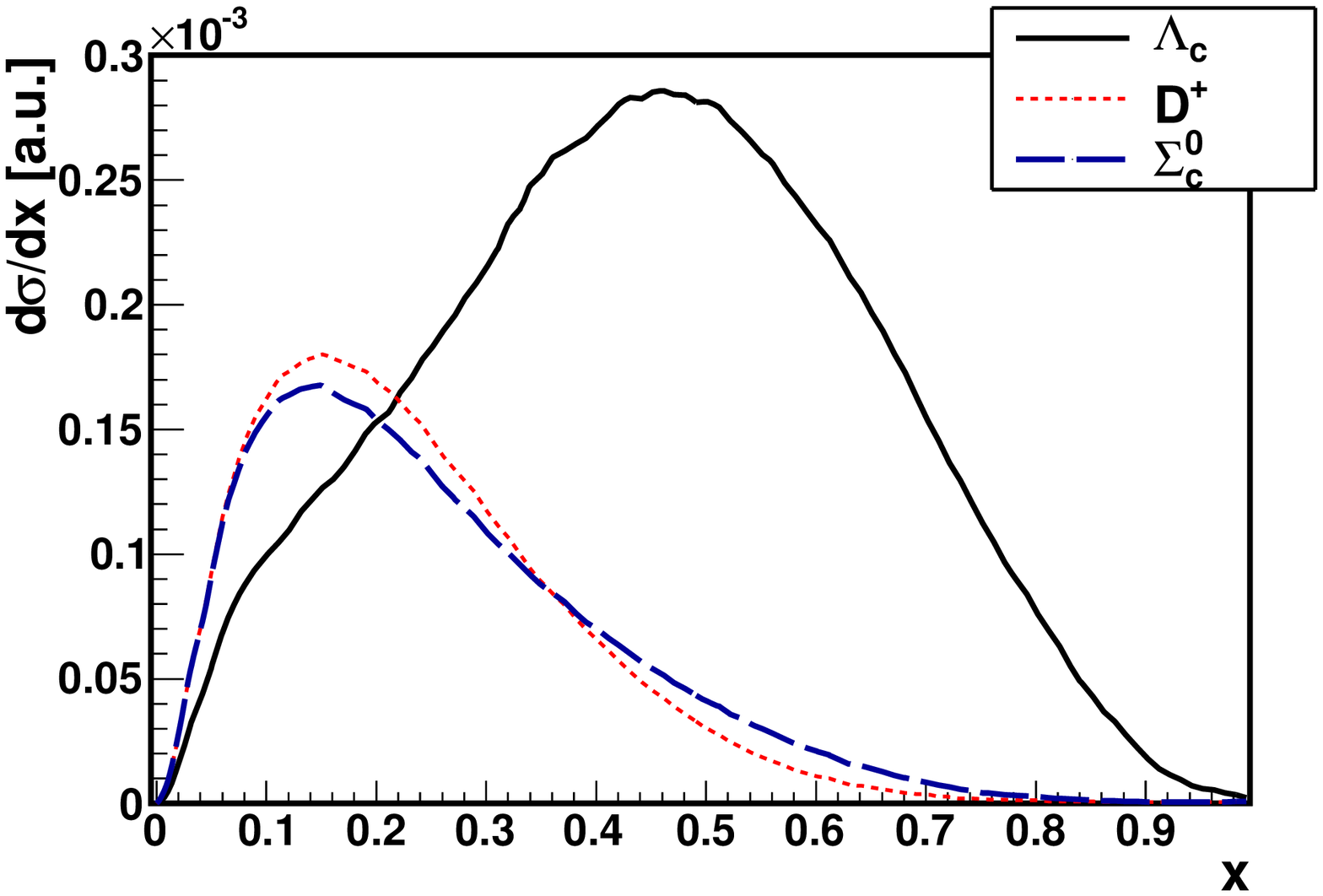}
   \includegraphics[width=0.49\textwidth]{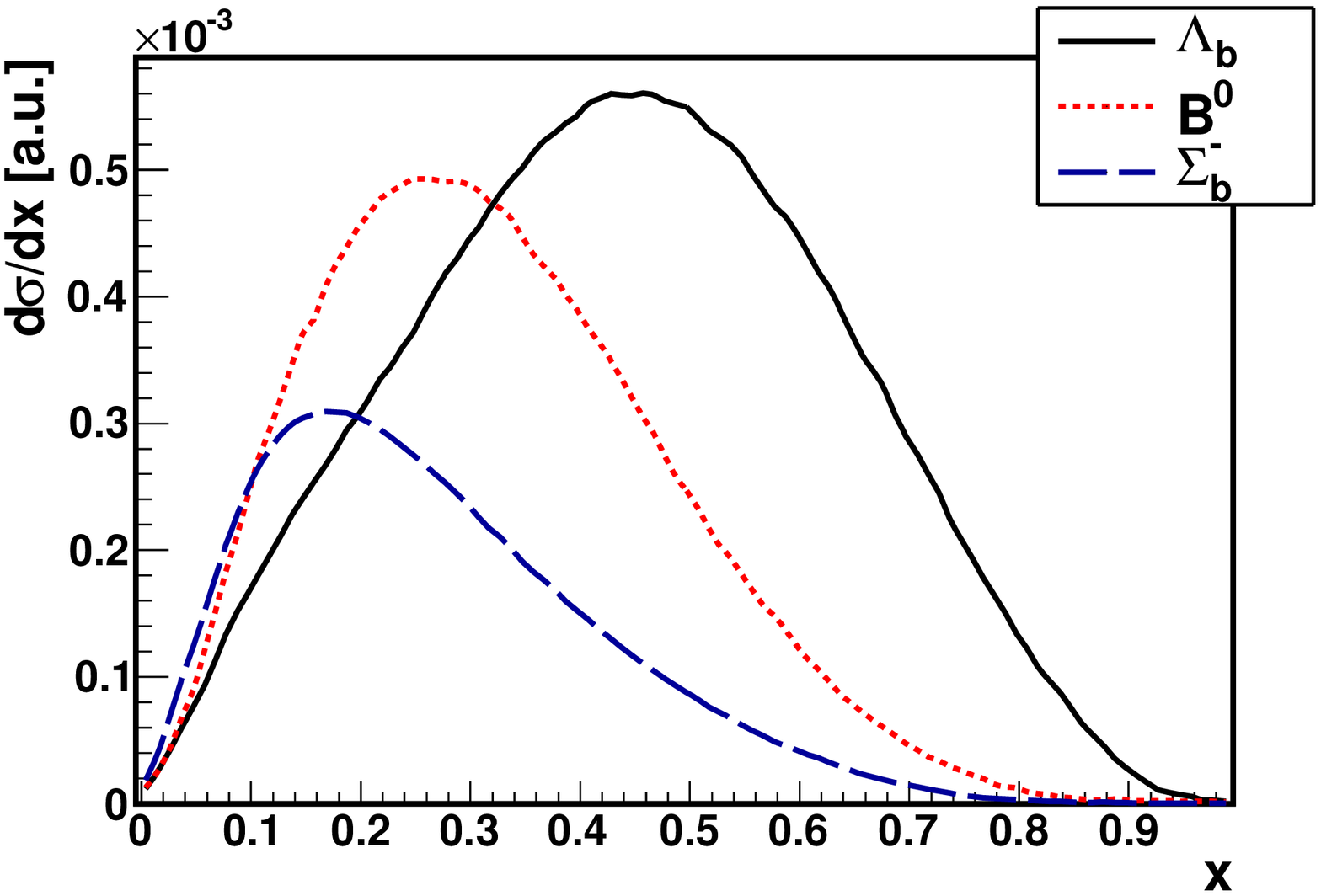}
  \end{center}
  \caption{Distribution of the fraction of primary energy in the Intrinsic Quark production model 
  for some charmed (left) and bottom (right) hadrons.}
  \label{fig:xfIQ} 

\end{figure}

At leading order in QCD, heavy quarks are produced by the processes $q\bar{q}\rightarrow Q\bar{Q}$ and $gg\rightarrow Q\bar{Q}$
(\figref{fig:xfCGC}, right). When these heavy quarks arise from fluctuations
of the initial state, its wave function can be represented as a superposition of Fock state fluctuations:
\begin{equation}
 |h> \,=\,  c_0|n_{v}> +\,\,c_1|n_v g> +\,\,c_2|n_v q\bar{q}> +\,\,c_3|n_v Q\bar{Q}>...
\end{equation} 
where $|n_v>$ is the hadron ground state, composed only by its valence quarks.
When the projectile scatters in the target the coherence of the Fock 
components is broken and the fluctuations can hadronize, either
with sea quarks or with spectator valence quarks. The latter mechanism is called Coalescence. 
For instance, the production of $\Lambda_{c}^+$ in p-N collisions 
comes from the fluctuations of the Fock state of the proton to $|uudc\bar{c} >$. To obtain
a $\Lambda_{c}^-$ in the same collision a fluctuation to $|uudu\bar{u}d\bar{d}c\bar{c} >$ would be
required. Thus, since the probability of a five quarks state is larger than that of a 9 quarks
state, $\Lambda_{c}^+$ production is favored over $\Lambda_{c}^-$ in proton reactions. 
The co-moving heavy and valence quarks have the same rapidity in these states but the larger mass
of the heavy quarks implies they carry most of the projectile momentum. Heavy hadrons
formed from these states can have a large longitudinal momentum  and carry a large fraction of
the primary energy \cite{ICofProton}, which is crucial for their
propagation. The differential energy fraction distribution for some charmed and bottom hadrons can be
seen in \figref{fig:xfIQ}.

This is the second model for heavy quark production we will implement. A detailed description of this 
model along with theoretical expressions for the differential cross-sections can be found in \cite{leadC,ICVogt,coalescence}
for intrinsic charm production and \cite{BottomVogt} for intrinsic bottom production.

\subsection{Physics of heavy hadron propagation}
\label{sec:propagation}

Charmed and bottom hadrons produced at accelerators are short-lived
particles and decay before interacting. In accelerator experiments
their presence is signalled through the detection of their final
state products. To measure, for instance, $\Lambda_c$-p interaction cross-sections, or the elasticity in
B-$\pi$ interactions, one would need to be able to produce beams of these particles above their critical
energies($\gtrsim10^{16}$ eV). Acceleration to those energies is so far out
of reach from a technological point of view. However, in collisions of
cosmic rays with momenta in the range of the EeV it is possible to
creat heavy hadrons with energies above their critical one. In this section we 
explain how we treat cross-sections, interactions lengths and
elasticities in this ultra-high energy regime.

\subsubsection{Interaction cross-sections and interaction lengths}
\label{subsec:propagation:cross-sections}

\begin{figure}[!t]
\begin{center}
  \includegraphics[width=0.49\textwidth]{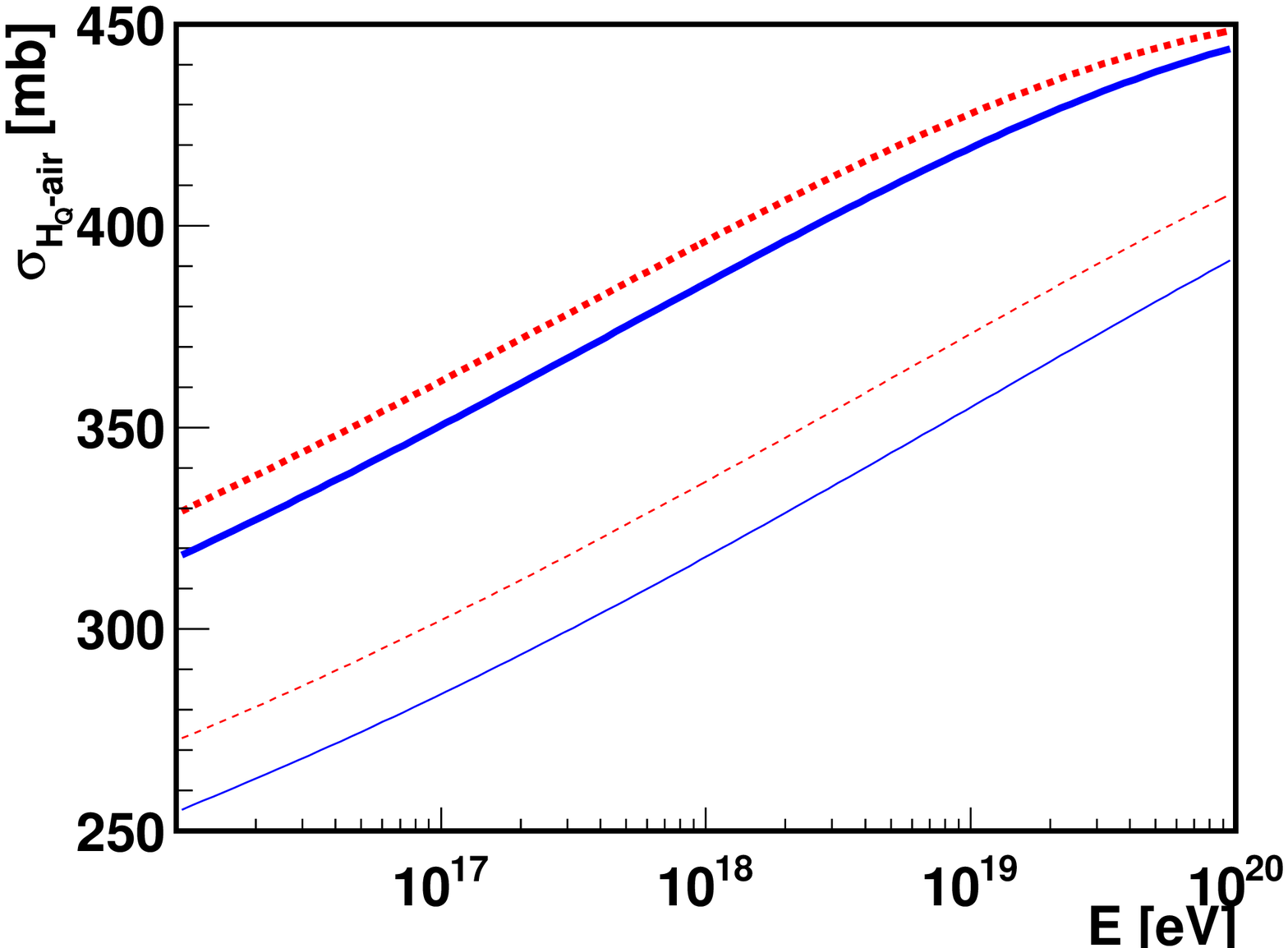} 
  \includegraphics[width=0.49\textwidth]{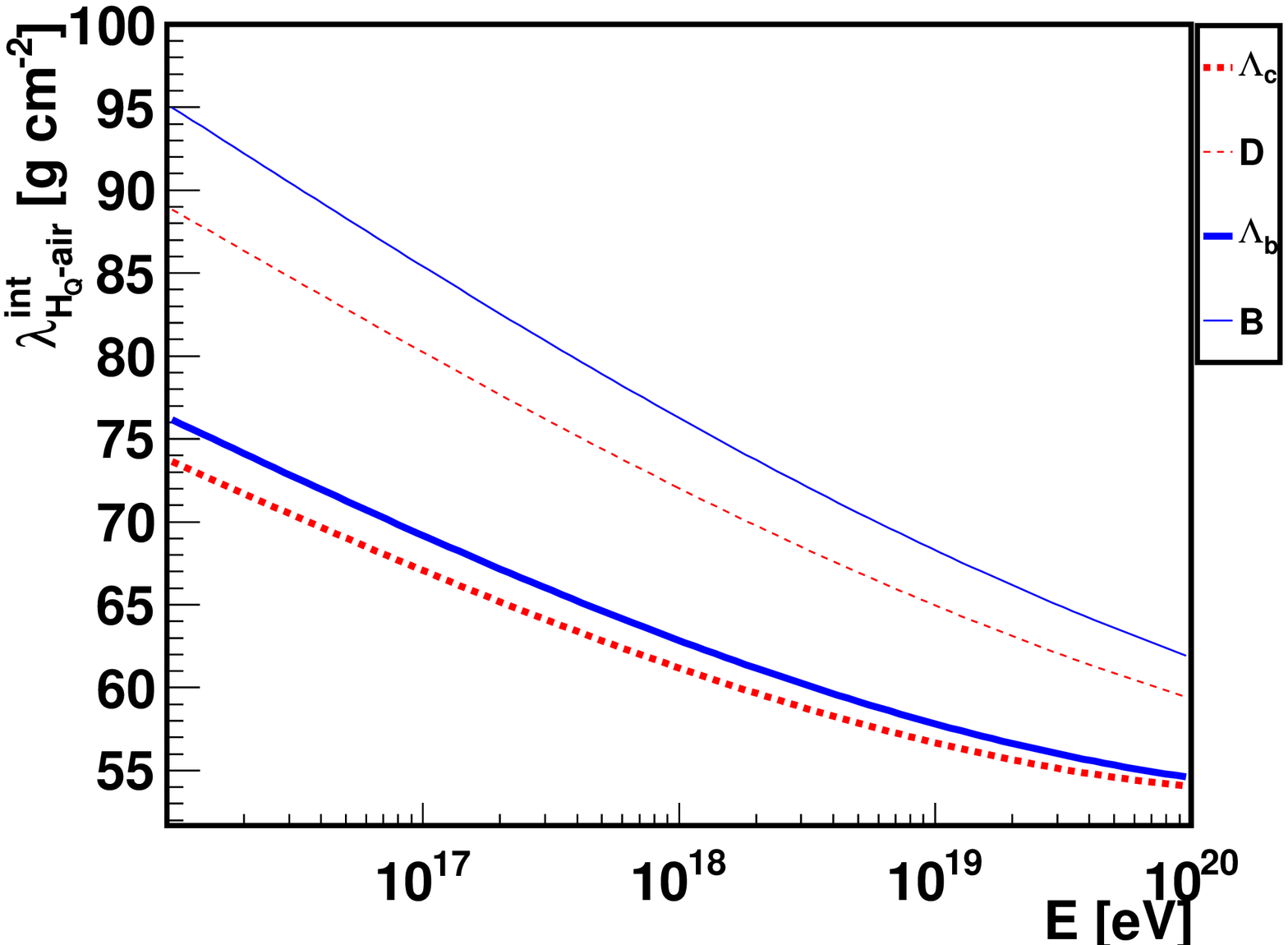} 
\end{center}
\caption{Left: Heavy hadron-air interaction cross-sections. Right: Interaction length in air. Thick and thin dashed lines 
for $\Lambda_c$ and D. Thick and thin solid lines for $\Lambda_b$ and B, respectively.}
\label{fig:crossint}
\end{figure}

Following the procedure explained in \cite{bottom_jcap} we obtain the inelastic cross-sections for
$\Lambda_c$, D, $\Lambda_b$ and B in collisions with protons at rest,
in an energy range going from $10^{16}$ eV
to $10^{20}$ eV. To scale these cross-section to collisions with air nuclei 
we apply the following prescription used in CORSIKA\footnote{The parameterization is inside the subroutine \textbf{BOX2}.}.
Let $\sigma^{H-p}$ be the hadron-proton cross-section.
Then, the hadron-air cross section is obtained as:
\begin{equation}
 \sigma^{H-air}~\textrm{[mb]}  =  (1-4\sigma_{45}^2)\cdot p_0 + \sigma_{45}(2\sigma_{45}-1)\cdot p_1 + \sigma_{45}(2\sigma_{45}+1)\cdot p_2
\end{equation} 
where 
\begin{eqnarray*}
\sigma_{45} & = & (\sigma^{H-p}~\textrm{[mb]} -45~\textrm{mb})/30 \nonumber \\
p_0 & = & 309.4268\, \textrm{mb} \nonumber \\
p_1 & = & 245.0771\, \textrm{mb} \nonumber \\
p_2 & = & 361.8057\, \textrm{mb}
\end{eqnarray*}  
From these cross-sections we can obtain the associated mean interaction length as 
\begin{equation}
\left< \lambda_{int} \right> = \left\langle m_{air} \right\rangle /\sigma^{H-air}
\label{eq:int_length}
\end{equation}
In \figref{fig:crossint} we can see the resulting cross-sections (left)
and interaction lengths (right) for heavy hadrons above $10^{16}$ eV. We use $\Lambda_c$ and $\Lambda_b$ cross-sections as representative
of all charmed and bottom baryons, respectively. In the same way, D and B cross-sections are used
for all charmed and bottom mesons. We adopt this criterion because, during their propagation,
heavier baryons and mesons transform into these lighter states during the first steps of the shower
development. For instance, $\Sigma_c$ ($\Sigma_b$) states rapidly decay 
into $\Lambda_c$ ($\Lambda_b$), which continue propagating
in the atmosphere.

\subsubsection{Interaction model}
\label{subsec:propagation:interaction}

The realistic implementation of heavy hadron propagation in EAS needs, apart from the values of cross-sections
and interaction lengths, the elasticity distributions of their interactions. 
We use the models described in \cite{bottom_jcap}
for the propagation of charmed and bottom hadrons. We analyze the collisions of very energetic D (B) mesons 
and $\Lambda_c$ ($\Lambda_b$) baryons with protons at rest. Using PYTHIA, diffractive and 
partonic collisions of any q$^2$ are considered to generate the corresponding elasticity distributions.

During their propagation in the atmosphere, heavy hadrons interact with air nuclei, and not with
protons. PYTHIA deals with hadron-nucleon collisions based on the Lund string model, but so far there is no agreement
on how hadron-nucleus collisions should be treated within this model. We will use the method described in \cite{sibyll_string}, 
which is the approach SIBYLL takes.

After a hadron-nucleon collision we find a leading hadron, carrying the largest
fraction of the primary energy, and a series of secondary particles, sharing
the rest of the energy. In a simplistic setting we could picture a hadron-nucleus
interaction as a series of independent hadron-nucleon collisions with every nucleon
composing the nucleus. As a result, the number of low energy secondaries produced would rise, increasing with each 
consecutive collision. Even though this is the behavior we would expect, the underlying assumptions are not correct. First, the time scales 
of the projectile traversing the nucleus and that of  
the recombination of partons inside the proton are fairly different, the former
being much shorter. Thus, there is no time for the proton to recombine
and suffer a second hadron-nucleon collision before it exits the nucleus. In addition, when a hadron 
collides with an air nucleus, not every nucleon within will participate in the interaction. 

To compute the number of nucleons participating in the interaction, N$_A$, 
we use the FORTRAN routine NUCOGE, where the probability of an inelastic
hadron-nucleon hit is determined by the choice of the hadron-nucleon overlap function.
A detailed explanation of how the program works can be found in \cite{NUCOGE}.
The next step is deciding the nature of the hadron-nucleus interaction, either diffractive
or partonic. The probability of a diffractive interaction, p$_{diff}$, is different for each projectile. 
In the case of charmed hadrons it is 0.30 for $\Lambda_c$ and 0.32 for
D. Turning to bottom hadrons, the values are 0.26 for 
$\Lambda_b$ and 0.29 for B. Let N$_D$ be the number of nucleons interacting diffractively. 
We will consider that the interaction is diffractive if, and only if, the
N$_A$ nucleons interact diffractively (N$_D=\, $N$_A$). 
If N$_D <\, $N$_A$, we consider that the collision is non-diffractive with N$_W =\, $N$_A - $N$_D$
participating nucleons. To treat the inelastic interaction of a heavy hadron, with energy $E_H$,
with an air nucleus we use the following prescription:
\begin{itemize} 
 \item First, from the N$_W$ participating nucleons, all but one are split 
  in quark-diquark pairs, i.e we have N$_W-1$ pairs and one unbroken nucleon.
 \item Then, N$_W-1$ quark-antiquark pairs are generated in the projectile, with total energy $E_{q\bar{q}}$.
 \item The partonic interaction occurs between the projectile, with 
  energy $E_H-E_{q\bar{q}}$, and the nucleon in the target that remains unaltered.
 \item The quark-antiquark pairs are matched with the quark-diquark pairs and hadronize.
\end{itemize}
Both the partonic interaction and the hadronization are performed by PYTHIA. As a final state, 
we find the particles resulting from the hard interaction plus
all the particles coming from the hadronization of the N$_W-1$ pairs formed. 

\begin{figure}[!t]
\subfigure[$\mathbf{\pi}$ \textbf{multiplicity}]{
  \includegraphics[width=0.49\textwidth]{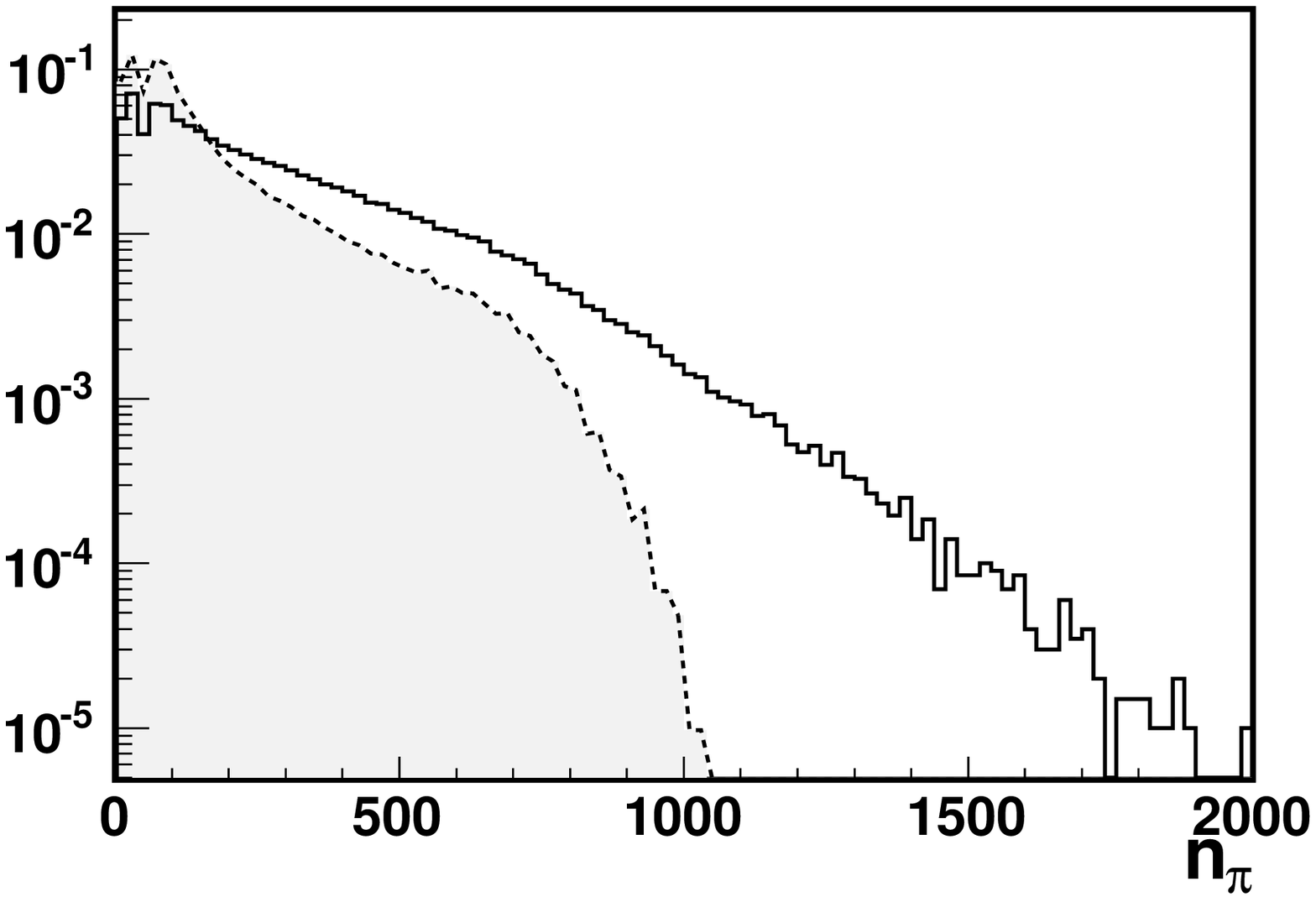}
\label{fig:PToAirA}
}
\subfigure[\textbf{K multiplicity}]{
  \includegraphics[width=0.49\textwidth]{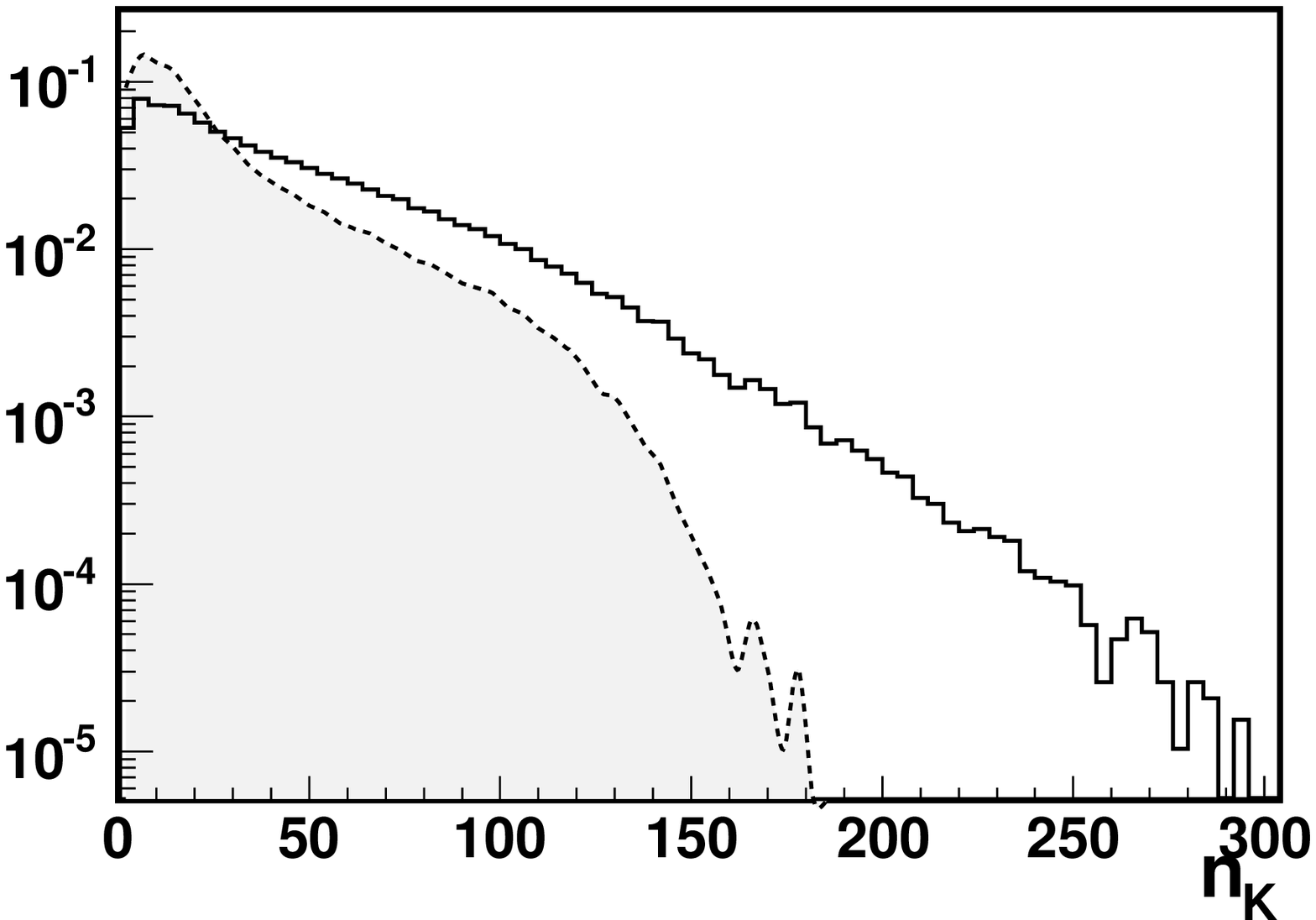}
\label{fig:PToAirB}
}
\subfigure[$\mathbf{\pi}$ \textbf{mean elasticity}]{
  \includegraphics[width=0.49\textwidth]{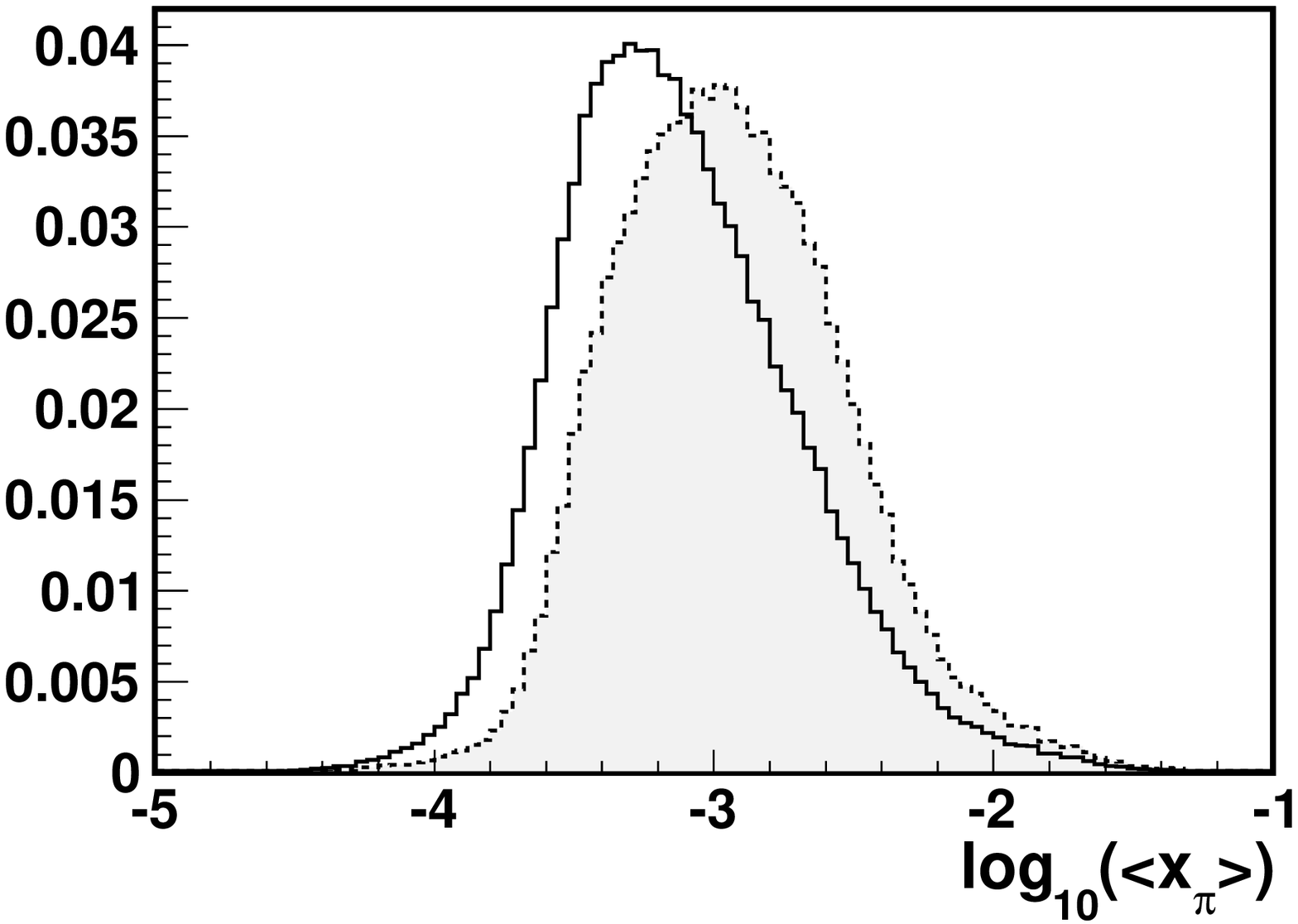}
\label{fig:PToAirC}
}
\subfigure[\textbf{K mean elasticity}]{
  \includegraphics[width=0.49\textwidth]{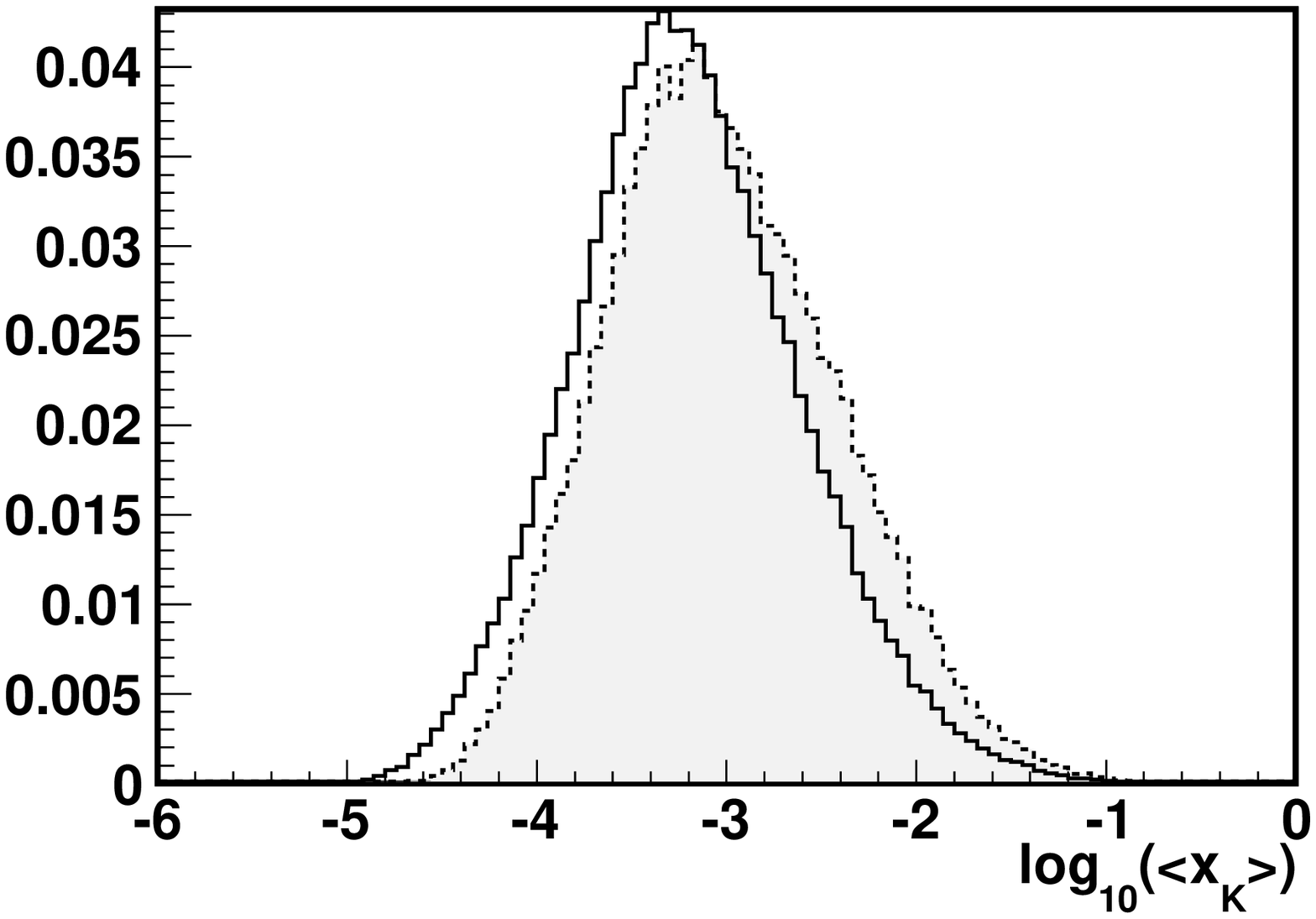}
\label{fig:PToAirD}
}

 \caption{Meson multiplicities and average elasticities 
for $\Lambda_b$-proton collisions (dashed lines) and
$\Lambda_b$-nucleus scattering (solid lines).}
\label{fig:PToAir}
\end{figure}

The main effect of the transition from hadron-nucleon to hadron-nucleus collisions is
increasing the multiplicity of produced particles, and decreasing their elasticity.
In \figref{fig:PToAir} we show the effect of the transition from hadron-proton (dashed line) to
hadron-air (solid line) collisions in case a $\Lambda_b$ is used as
the probing projectile. All distributions are scaled to the same integral. 
In \figref{fig:PToAirA} and \figref{fig:PToAirB} we observe that the multiplicity of pions and kaons 
is larger in collisions with air. At the same time (\figref{fig:PToAirC} and \figref{fig:PToAirD})
the energy transferred to the pion component barely rises and thus the average elasticity
per secondary particle is smaller (by a factor 1.5): 
a larger number of particles is sharing roughly the same amount of energy.

\begin{figure}[!t]
\subfigure[\large{$\mathbf{\Lambda_c}$}]{
\includegraphics[width=0.49\textwidth]{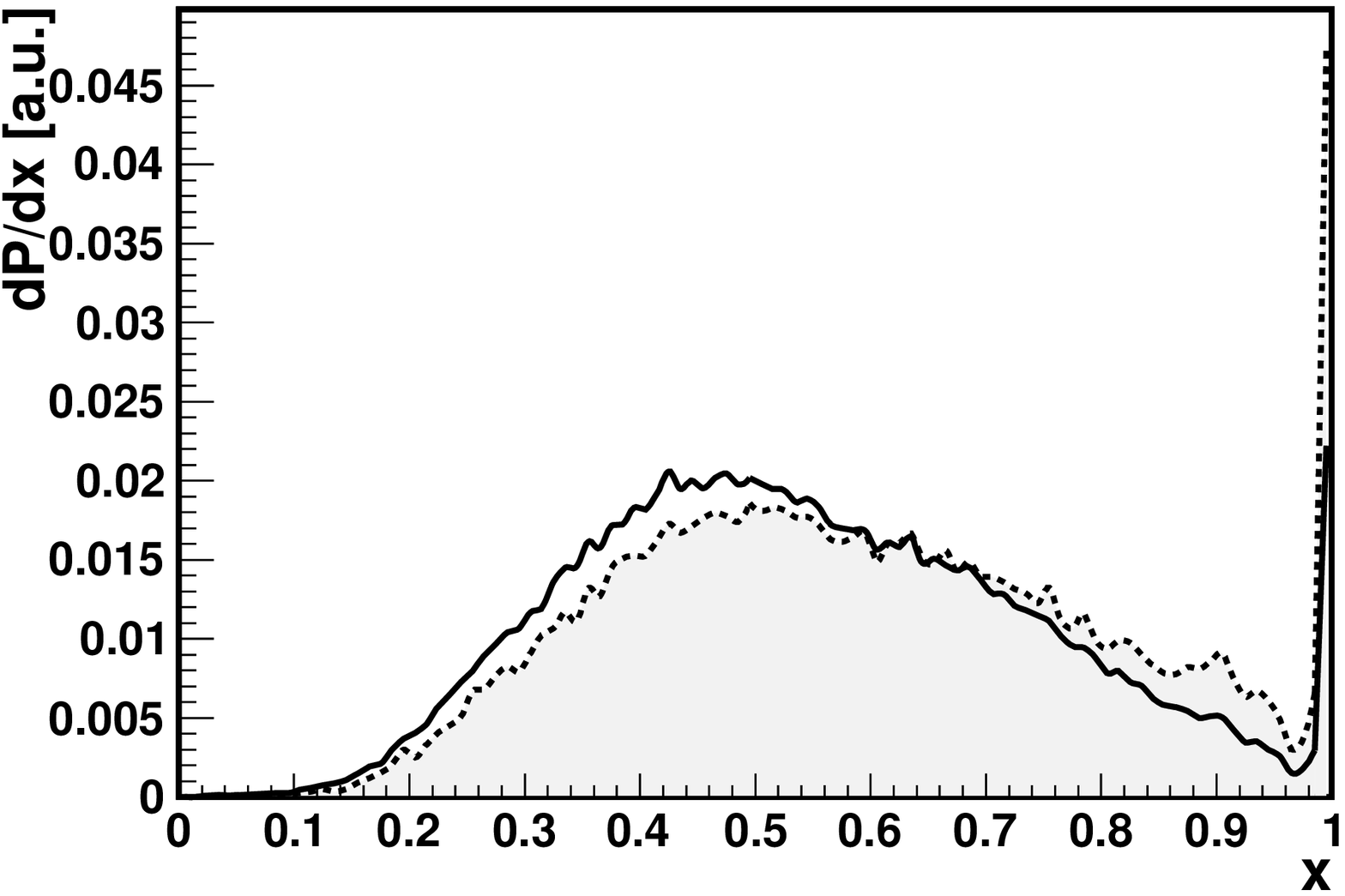}
\label{fig:4plotsA}
}
\subfigure[\large{$\mathbf{D^+}$}]{
\includegraphics[width=0.49\textwidth]{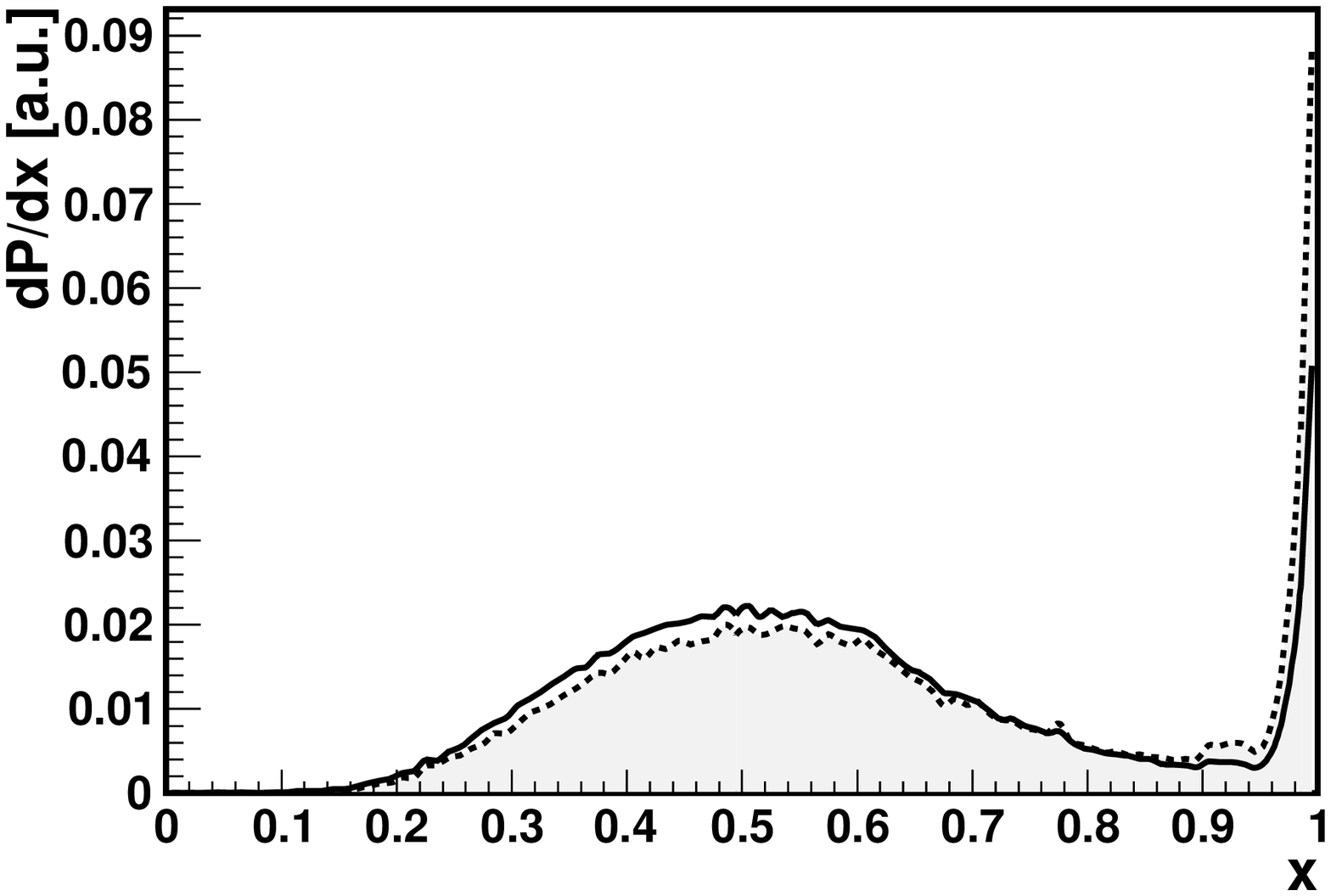}
\label{fig:4plotsB}
}
\subfigure[\large{$\mathbf{\Lambda_b}$}]{
\includegraphics[width=0.49\textwidth]{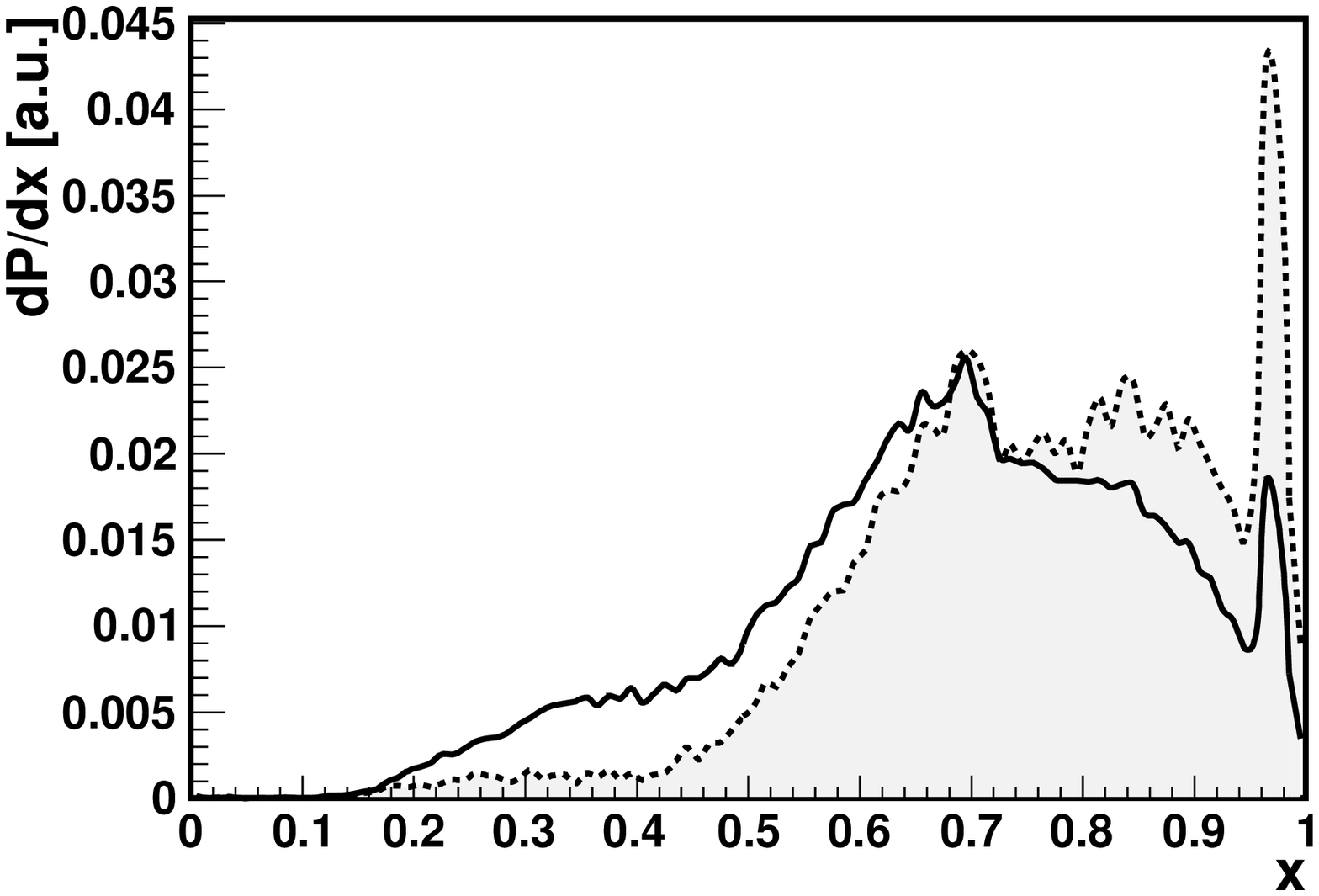}
\label{fig:4plotsC}
}
\subfigure[\large{$\mathbf{B^+}$}]{
\includegraphics[width=0.49\textwidth]{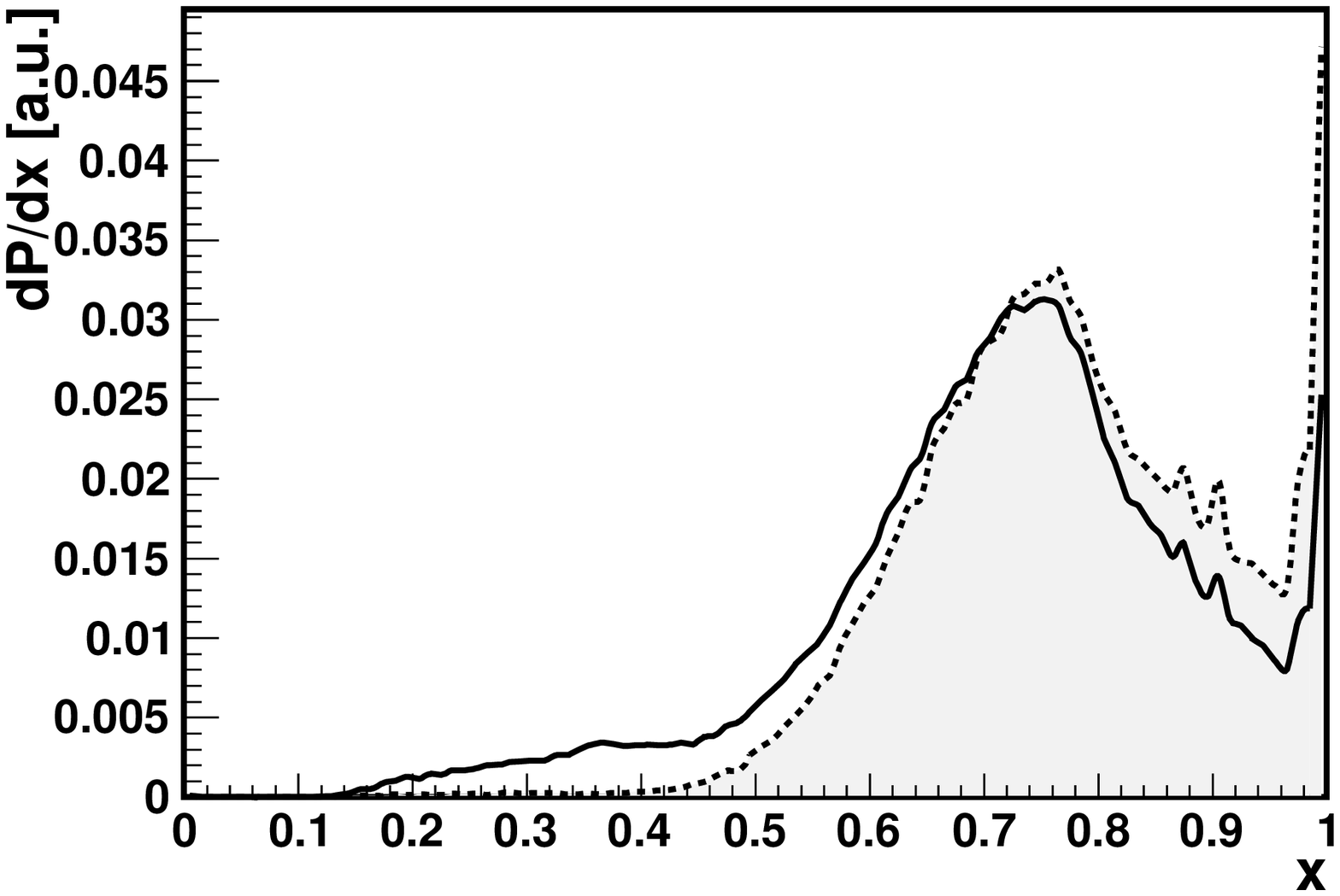}
\label{fig:4plotsD}
}
\caption{Elasticity distributions of the leading hadron after
  collisions off protons (dashed lines) or air (solid lines), for four different
 projectiles.}
\label{fig:4plots}
\end{figure}

Interacting with air nuclei rather than with protons also affects the leading hadron. In \figref{fig:4plots} 
we can see the elasticity distributions of the leading charmed and
bottom hadrons after collisions with protons compared to those where
the collisions take place off air nuclei (solid lines).

\begin{table}[!t] \large 
	\caption{Mean elasticity values for the collisions of heavy hadrons with protons and air.}
	\begin{center}
		\begin{tabular*}{\textwidth}{@{\extracolsep{\fill}} c c c c c }
		\hline
       	            & $\Lambda_c$ & $D^{+}$    & $\Lambda_b$ & $B^{+}$ \\ \hline
		p   & 0.62        & 0.65 & 0.72        & 0.75 \\ \hline
		Air & 0.56        & 0.59 & 0.68        & 0.72 \\ \hline
		\end{tabular*}
                \label{tab:mean_elas}
	\end{center}
\end{table}

All distributions are scaled to the same integral. Regarding the comparison between charmed
and bottom hadrons collisions, we find that the latter are, on average, more elastic than the former. 
That is because the heavy quark composing the hadron behaves as a spectator most of the
 time, losing a small amount of its energy in the interaction. In addition, the fraction 
of the heavy hadron energy carried by the heavy quark is proportional to the heavy quarks mass.

Collisions with air nuclei are also more inelastic than collisions with protons. In hadron-nucleus collisions
there might be more than one nucleon involved. Interacting with more nucleons increases the number of soft secondaries,
and decreases the energy fraction carried away by the leading particle. The mean elasticity values for collisions with protons
and air can be found in \tabref{tab:mean_elas}.

\section{Simulation chain and code implementation}
\label{sec:simchain}

In this section we describe the different steps involved in the Monte
Carlo simulation of heavy hadrons in EAS and reference the 
subroutines modified or newly written. The reader can find a summary of the modifications made to the CORSIKA source code 
in \ref{sec:code:code}. As we mentioned in section \ref{sec:introduction}, Monte Carlo simulators
 do not handle charmed hadrons production or, if they do,
they are not propagated. In the case of bottom hadrons, they are not
even included in the list of particles recognized by the programs. Our
goal is to implement these particles, such that they are eligible candidates
for production and propagation.

The simulation chain consists of several steps. Initially, we simulate the
primary particle first interaction, choosing whether charmed hadrons, bottom hadrons or none of them
are produced in the collision.  The propagation of heavy hadrons across the atmosphere 
takes place along with the rest of the shower, but according to the interaction model described in
section \ref{subsec:propagation:interaction}. The decay of both charmed and bottom hadrons is performed by PYTHIA. 
\subsection{First interaction}
\label{subsec:first}
\begin{figure}[!t]
\begin{center}
  \includegraphics[width=0.49\textwidth]{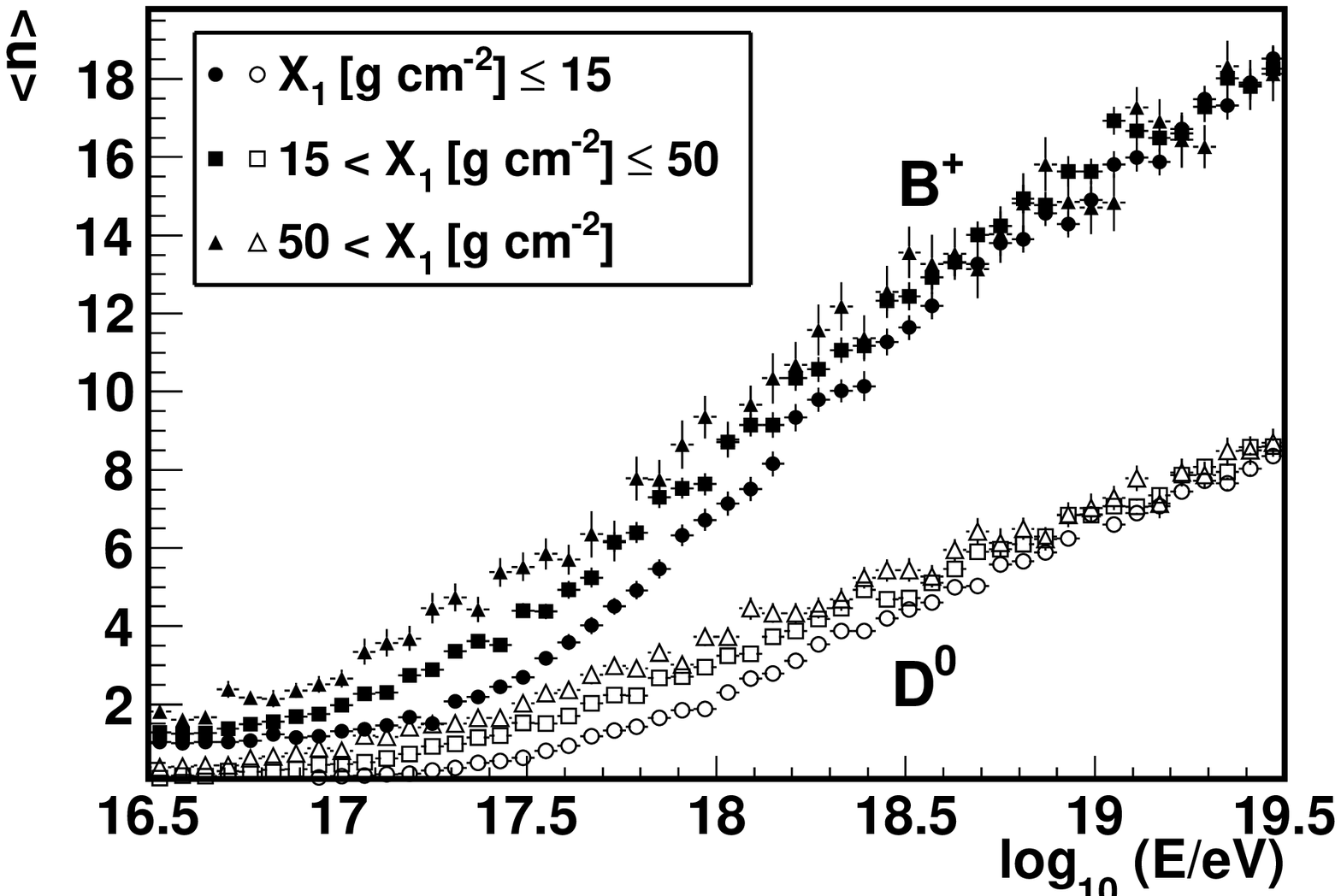} 
  \includegraphics[width=0.49\textwidth]{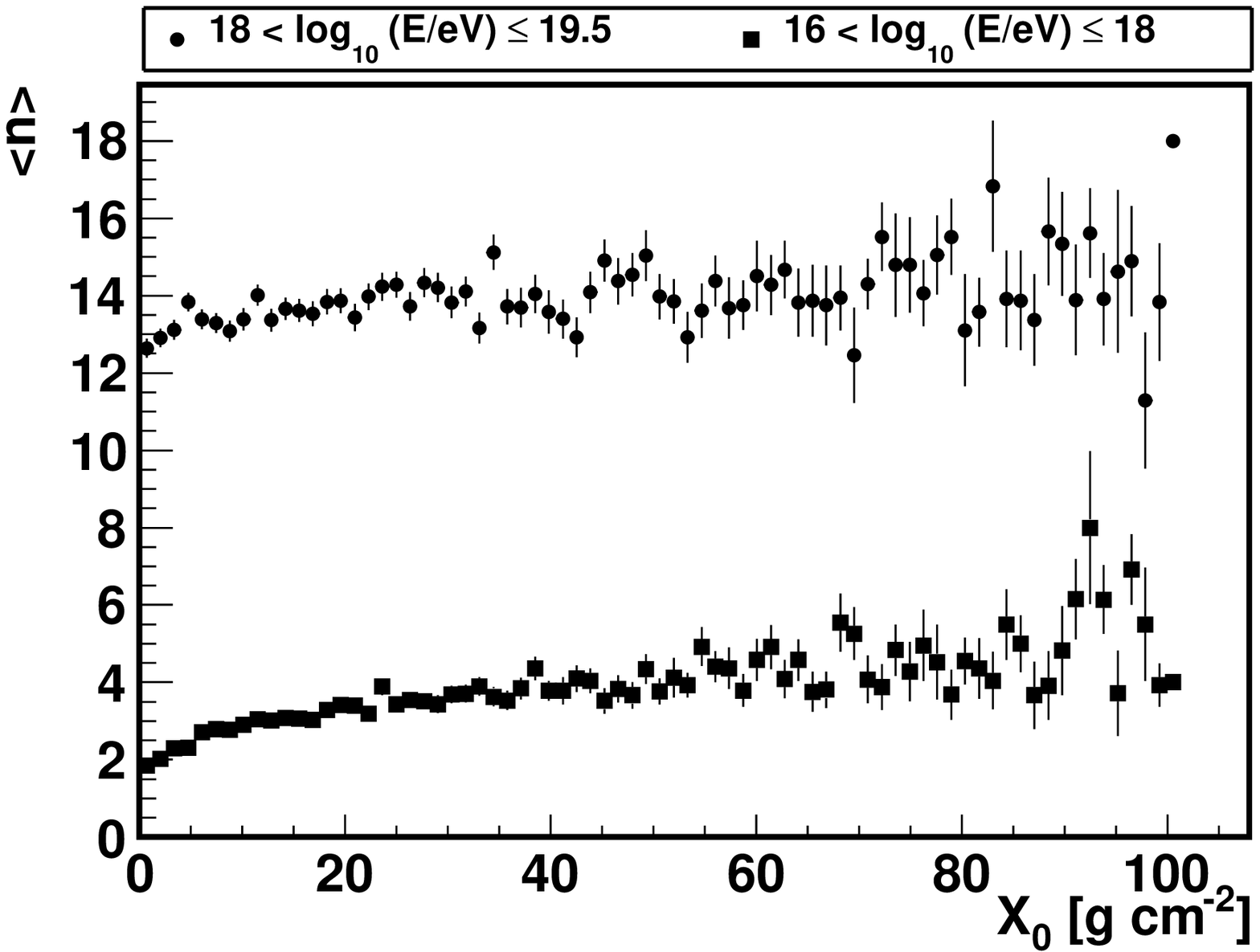} 
\end{center}
\caption{Left: Mean number of interactions, $<$n$>$ as function of energy, for different production depth bins.
Right: Mean number of interactions $<$n$>$ of a $B^+$ meson as function of depth, for different energy bins.}
\label{fig:meanN}
\end{figure}

Heavy quarks can be produced in any of the collisions taking place
along the shower development, provided the interaction is energetic
enough. However we restrict our interest only to heavy hadrons produced in the first
interaction of the primary particle with an atmospheric nucleus. 
Charmed and bottom hadrons produced in subsequent interactions are much less
energetic and therefore their influence in the longitudinal
development of the shower will be small. And, even though they could be produced
deeper in the atmosphere, it is the energy, and not the production depth, that rules the propagation.
To check this, we simulate $B^+$ and $D^0$ mesons with a uniform distribution in log$_{10}($E/eV$) \in \left[ 16.5,19.5 \right]$.
The production depth corresponds to the depth of the proton first interaction, distributed as
\begin{equation}
 P(X_{0};\lambda^{p-Air}_{int}) = \frac{1}{\lambda^{p-Air}_{int}} \exp(-X_{0}/\lambda^{p-Air}_{int})
\end{equation}
In \figref{fig:meanN} (left) we plot the mean number of interactions suffered by the $B^+$ and $D^0$ in the atmosphere before decaying 
 as a function of the initial meson energy, for different ranges of production depths.
In \figref{fig:meanN} (right) we can find the mean number of interactions suffered by the $B^+$, as a function of X$_0$, for different primary
energy ranges. The number of interactions suffered before decay
increases rapidly as the meson initial energy grows, almost
independently of $X_0$. Whereas for fixed energies, the number of interactions is roughly constant with growing production depth.

The subroutine \textbf{COLLIDE} produces the charmed or bottom hadrons
right after the primary proton first interaction. As we mentioned in section \ref{subsec:CGC} heavy hadrons in the Color Glass Condensate model 
are formed via fragmentation. Thus, all heavy hadrons are formed with equal probability.
In the case of Intrinsic Quark production, the proton develops a fluctuation to a 5- or 7-particle
state before the collision with an air nucleus. We will not consider higher fluctuations.
During the collision, each of the heavy quarks composing the fluctuation will independently	
hadronize, either by coalescence or fragmentation, with probability 50\%. However, the allowed
final states will not always be independent: when hadronization occurs by fragmentation any hadron
can be formed; upon hadronization by coalescence, the accessible states are limited by the quark
content of the fluctuation. For instance, if the 
proton develops a $|uudb\bar{b}>$ fluctuation, no bottom hadrons containing a $s$ quark can be formed
by coalescence. Thus, the distributions of the primary energy fraction that goes into different hadrons will
be different (see for example \figref{fig:xfIQ}).

\subsection{Propagation}
\label{subsec:propagation}

After their production, the heavy hadrons generated at the first interaction 
have to be propagated. This is the process largely neglected in air shower simulators.
As CORSIKA has been modified to recognize particles with bottom quarks, both charmed and
bottom hadrons can be propagated using the standard machinery built in CORSIKA.
During their propagation, these particles will interact with nuclei in the atmosphere
or will decay in flight. Whether any of these happens depends on the values
of the interaction and decay lengths. The mean interaction length in units of 
depth is given by Eq. \ref{eq:int_length}. The interaction lengths for 
charmed and bottom hadrons are plotted in \figref{fig:crossint} (right).
The mean decay length, i.e the mean distance a particle traverses before it decays,
in units of distance is given by:
\begin{equation}
 \left< \lambda_{dec} \right> = \frac{E c \tau}{m}
\end{equation}
 where $\tau$ is the particle mean life-time and $m$ its mass. In CORSIKA, the actual values for the
 interaction and decay lengths are sampled from the following exponential distributions:
 \begin{eqnarray}
  P(\lambda_{int};\left< \lambda_{int} \right> ) = \frac{1}{\left< \lambda_{int} \right>}  & \exp(-\lambda_{int}/\left< \lambda_{int} \right>) \\
  P(\lambda_{dec};\left< \lambda_{dec} \right> ) = \frac{1}{\left< \lambda_{dec} \right>}  & \exp(-\lambda_{dec}/\left< \lambda_{dec} \right>)
 \end{eqnarray}
If $\lambda'_{dec}<\lambda_{int}$, where $\lambda'_{dec}$ is the decay length expressed in depth units, the particle 
travels a distance $\lambda'_{dec}$ and decays. Else, the particle travels $\lambda_{int}$ before interacting
with an atmospheric nucleus. Energy losses during the particle
time-of-flight are treated by CORSIKA standard routines.

\subsection{Interaction}
\label{subsec:interaction}

As heavy hadrons cross the atmosphere, they will collide with atmospheric nuclei.
We treat the collisions according to the model described in section \ref{sec:physics}.
The new subroutine \textbf{HEPARIN} links with the PYTHIA routines
 that treat the interaction of heavy hadrons with air nuclei, instead
 of calling the high-energy hadronic model chosen during compilation. It calculates
 the number of interacting atmospheric nucleons using the function
 \textbf{NNY} and assigns whether the interaction is diffractive or
 partonic. After each collision numerous particles are generated, and
 usually  the particle containing the heavy
quark carries away the largest energy fraction. All the collision products
 are injected back to the CORSIKA stack using \textbf{PYTSTO} and
 tracked as any other particle that contributes to the shower
 development. 

\subsection{Decay}
\label{subsec:decay}

During their propagation in the atmosphere the heavy hadrons will lose
energy due to bremsstrahlung and ionization, but specially because of
their collisions with nuclei. The decrease in energy modifies the values
of both the interaction and decay lengths, rising the former and 
reducing the latter, and thus increasing the decay probability. At the same
time, the particle approaches ground and the atmosphere grows thicker, 
reducing the distance between interactions. The interplay of these effects
will determine where the decay occurs.

The decay of both charmed and bottom particles is performed within
CORSIKA. \textbf{BTTMDC}, a new subroutine, is called to treat the decay
of bottom hadrons. It is analogous to the already
existing \textbf{CHRMDC} CORSIKA routine.

\section{Effects on shower propagation}
\label{sec:effects}

We have used the modified CORSIKA package described in previous
sections to generate large statistical samples of showers where charm
or bottom quarks are produced in the first interaction. For each heavy
quark, we have generated a library that contains more than one hundred
thousand events. Our goal is to understand how 
the presence of a heavy hadron could affect
fundamental parameters of the cascade development like the shape of its
longitudinal profile, the number of particles reaching ground, the
position and amplitude of the shower maximum, etc.  Heavy hadrons propagating 
with an energy above their critical one will fly over long paths. After several
elastic interactions, we expect them to deposit their remaining energy
deep in the atmosphere and some might even reach ground. In the case
the heavy hadron carries a significant fraction of the energy of the
primary particle that created the shower, one naturally expects that 
both the size and the shape of the cascade will be altered. 
The larger the energy of the heavy component, the more accentuated these effects will be. 

\begin{figure}[!t]
  \includegraphics[width=0.49\textwidth]{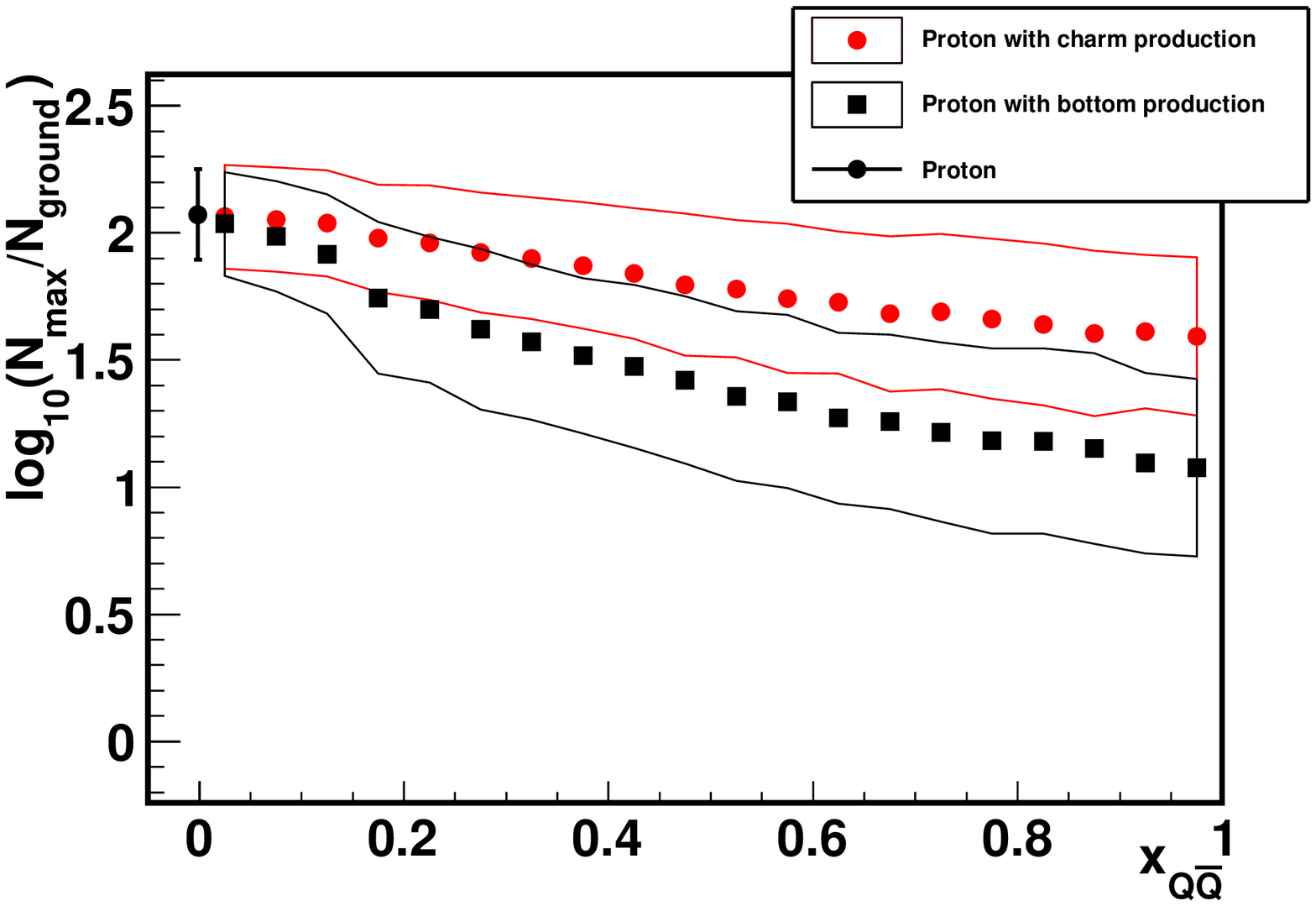}
  \includegraphics[width=0.49\textwidth]{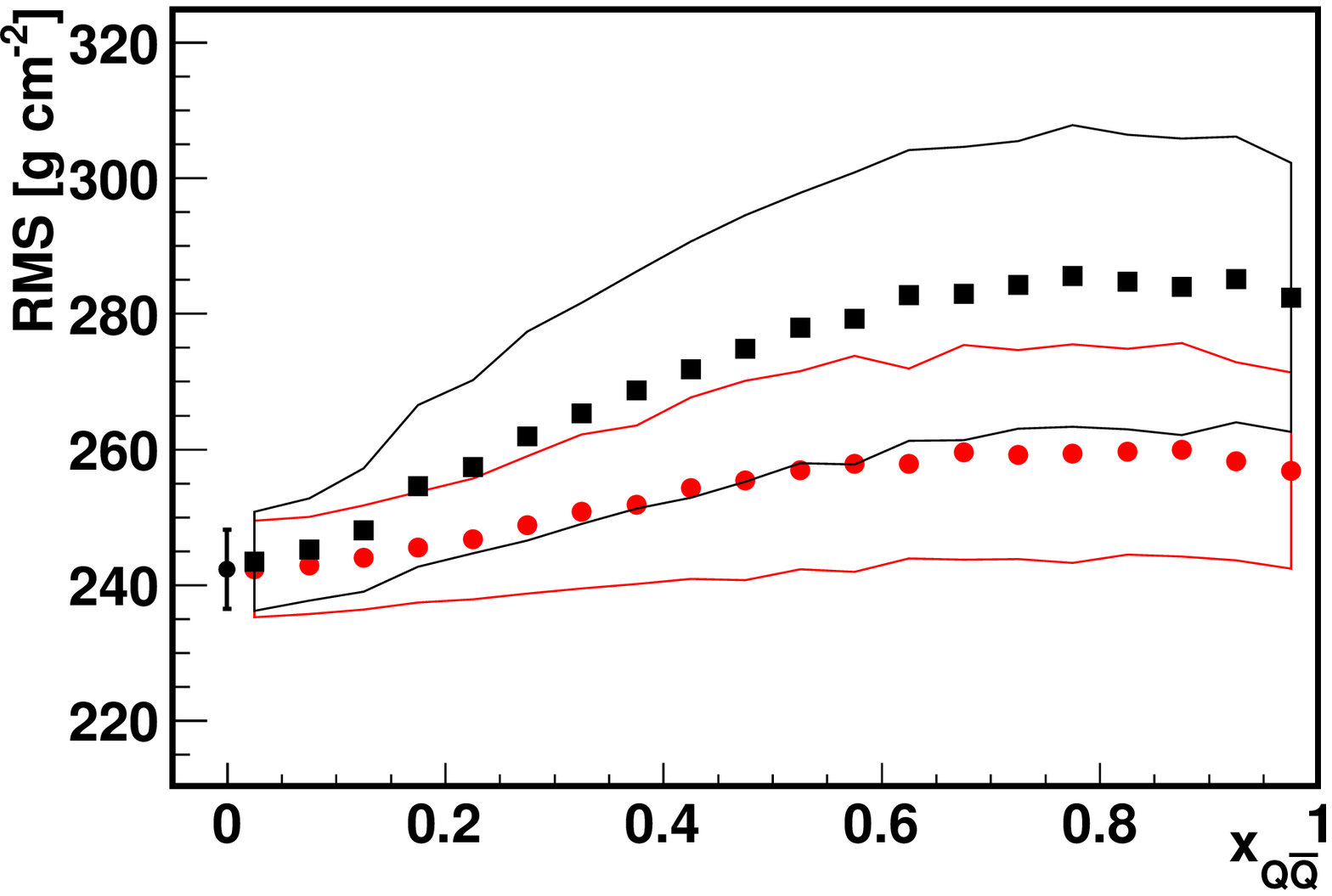}
 \caption{RMS (right) and ratio of the number of particles at shower
   maximum to the number of particles at ground (left) as a function
   of the fraction of energy carried by the heavy hadron. Open symbols
   correspond to showers with heavy hadron production on. Full symbols are the expected values for proton showers
   with no heavy quark production.} 
 \label{fig:MCNmaxNground}
\end{figure}

Let us analyze the case where the energy deposition in the shower is
shifted to larger depths. The position and the amplitude of the shower
maximum will be affected. The number of particles at maximum will
decrease, while at the same time the number of particles that reach ground will increase.
In \figref{fig:MCNmaxNground} (left) we plot the ratio of the number of particles at shower maximum 
with respect to the number of particles at ground. This ratio is plotted as
a function of the energy 
fraction carried away by the heavy hadrons produced in the first interaction. 
As this fraction grows, the ratio decreases, which means that a component of the shower is 
being displaced to larger depths. The bands show the $\pm 1 \sigma$
deviation. For comparison, we superimpose (solid symbol) the average value $\pm 1 \sigma$ deviation for
proton showers where heavy quark production has been turned off. 
For similar reasons shower profiles with a leading heavy hadron must be wider on average, with larger RMS.
This effect must increase with rising fractions of the energy carried away by the heavy hadrons.
As shown in \figref{fig:MCNmaxNground} (right) our hypothesis is
confirmed by simulations. The mean RMS of the showers increases with
the energy fraction transferred to the heavy quark. For reference, we plot the expected RMS value 
along with its $\pm1\sigma$ deviation for
proton showers where no heavy quark production is allowed.

\begin{figure}[!t]
 \begin{flushleft}
\subfigure[]{
\includegraphics[width=0.31\textwidth]{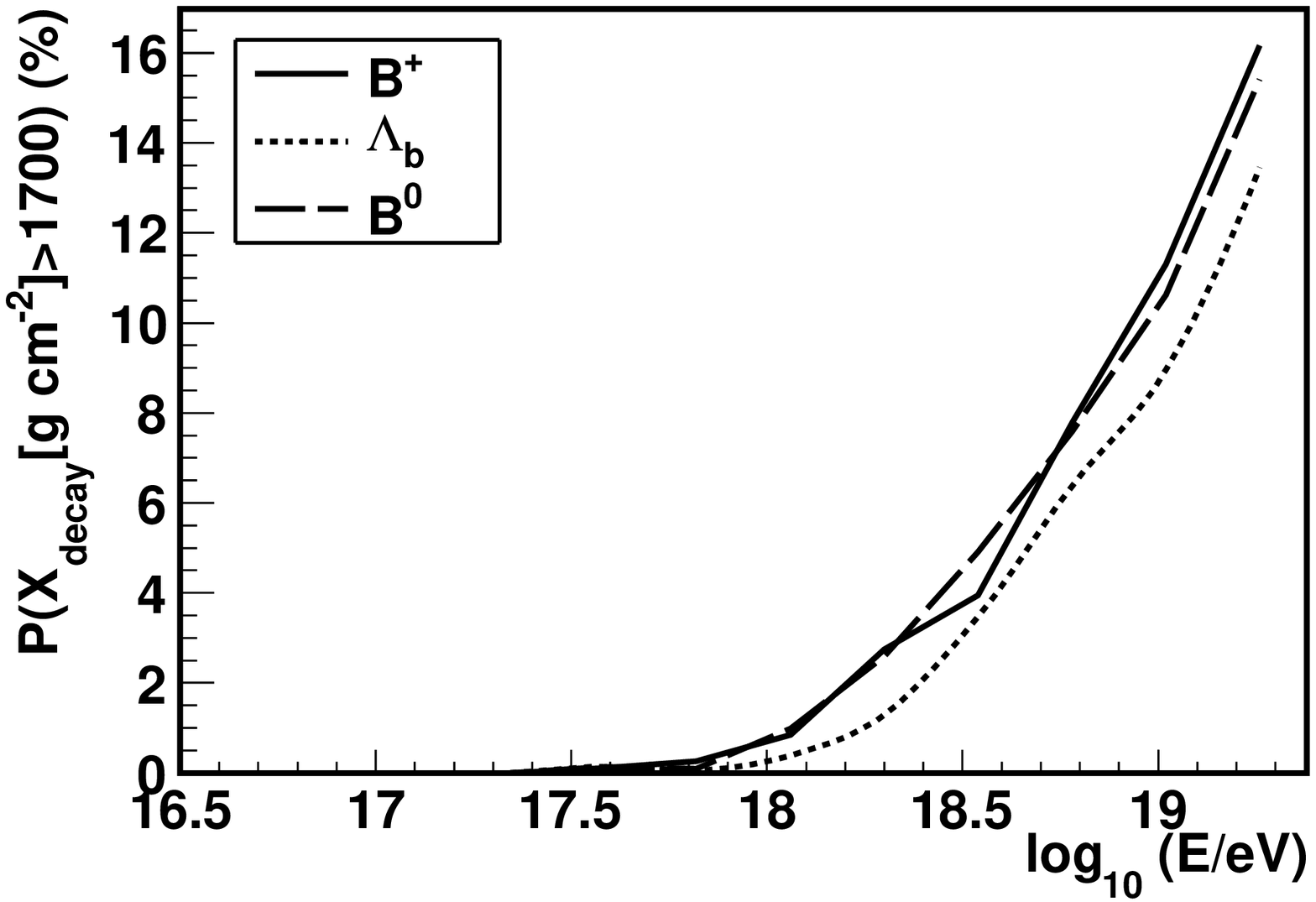}
\label{fig:decayA}
}
\subfigure[]{
 \includegraphics[width=0.31\textwidth]{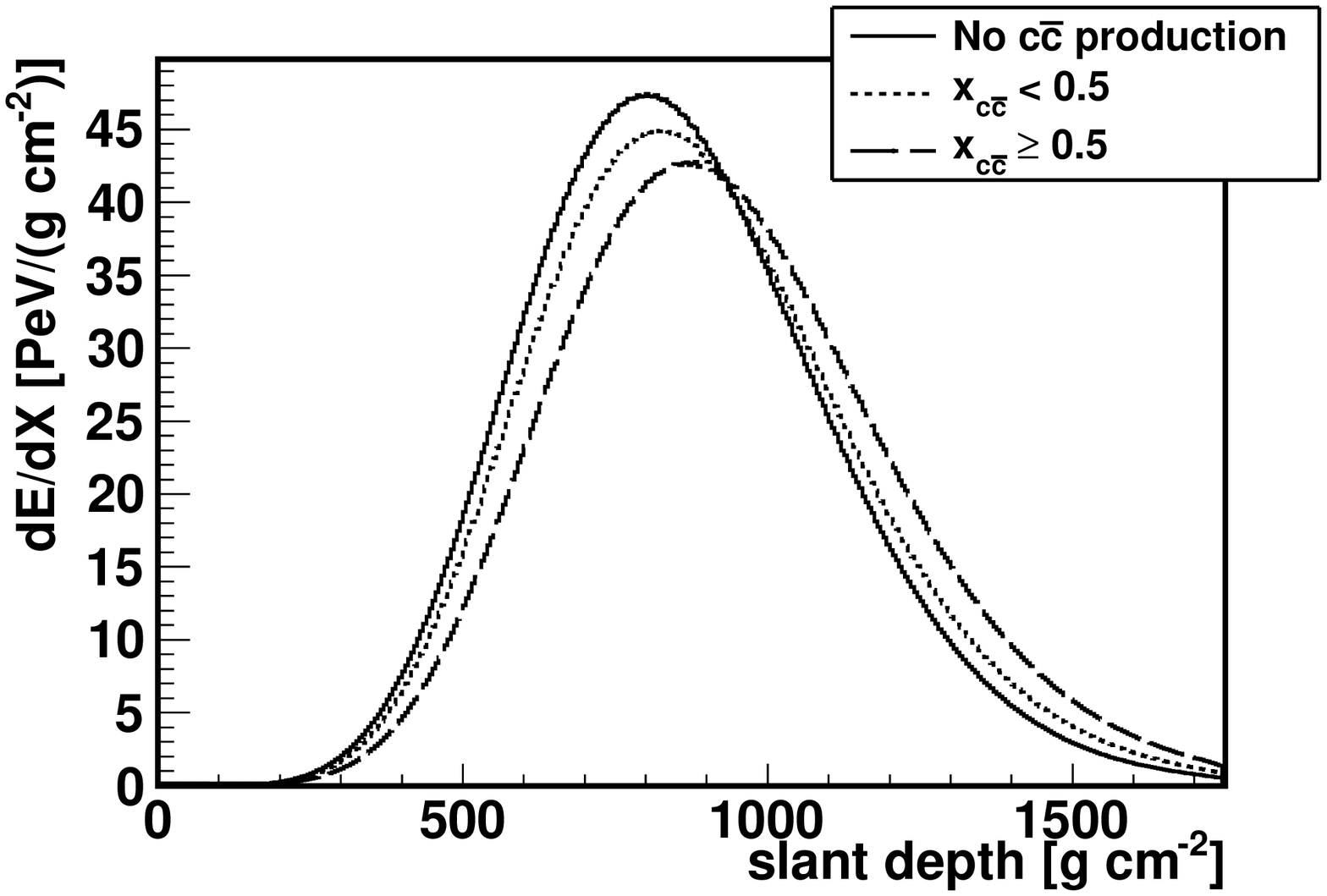}
\label{fig:decayB}
}
\subfigure[]{
 \includegraphics[width=0.31\textwidth]{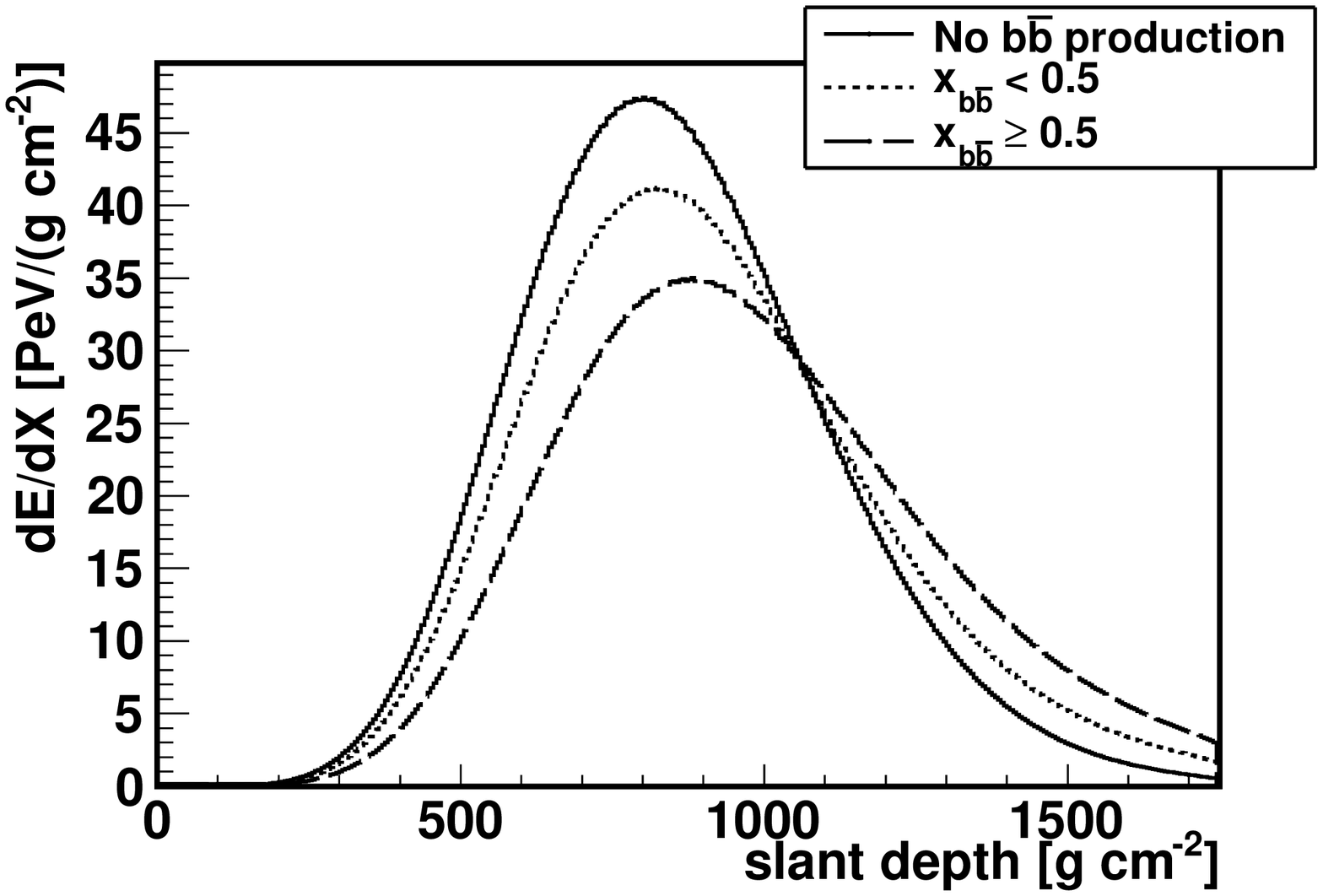}
\label{fig:decayC}
}
 \end{flushleft}
 \caption{(a): Probability of decaying above 1700 g cm$^{-2}$ for different bottom hadrons as a function of their initial energy.
(b) and (c): Mean $10^{19.5}$ eV shower profiles for proton showers
with no heavy quark production (solid line), with heavy hadrons 
carrying less than 0.5 of the primary energy (dashed line) and carrying more than 0.5 of the primary energy (long dashed line).
}
  \label{fig:decay}
\end{figure}

\begin{figure}[!t]
  \subfigure[\textbf{Proton showers}]{
    \includegraphics[width=0.49\textwidth]{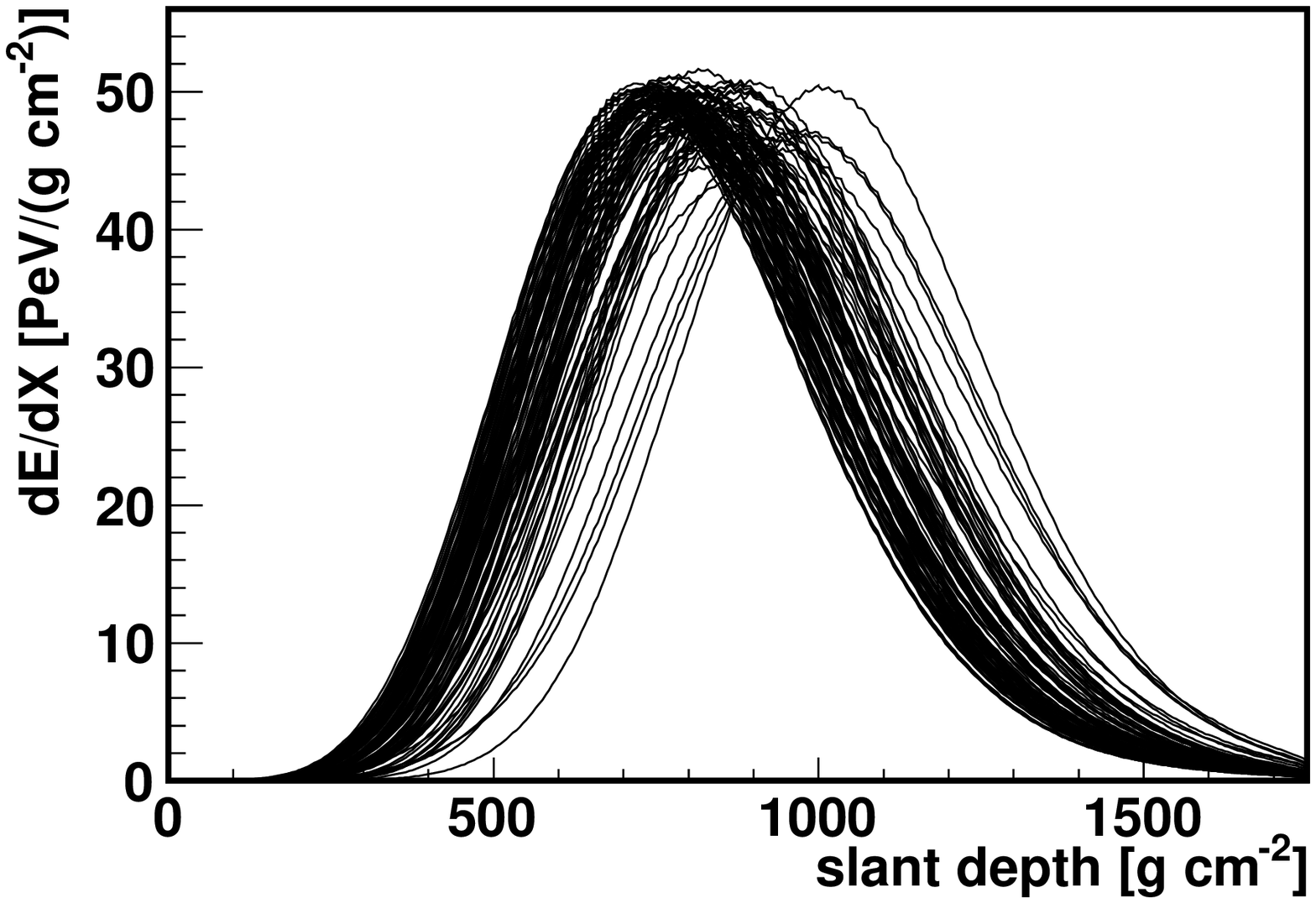}
    \label{fig:all_profiles_proton}
  }
  \subfigure[$\mathbf{x_{b\bar{b}}<0.5}$]{
    \includegraphics[width=0.49\textwidth]{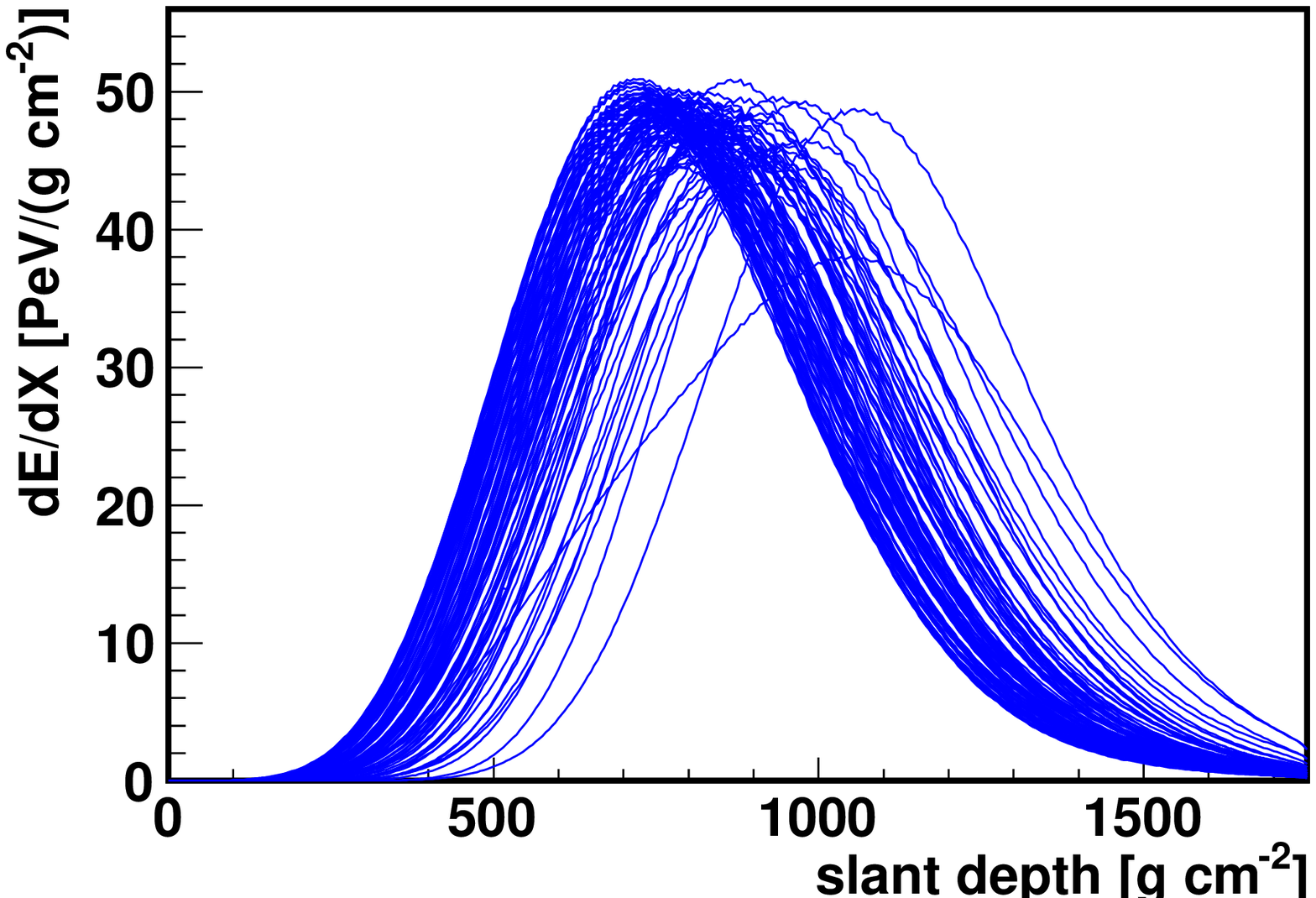}
    \label{fig:all_profilesB1}
  }
  \subfigure[$\mathbf{x_{b\bar{b}}>0.5}$]{
    \includegraphics[width=0.49\textwidth]{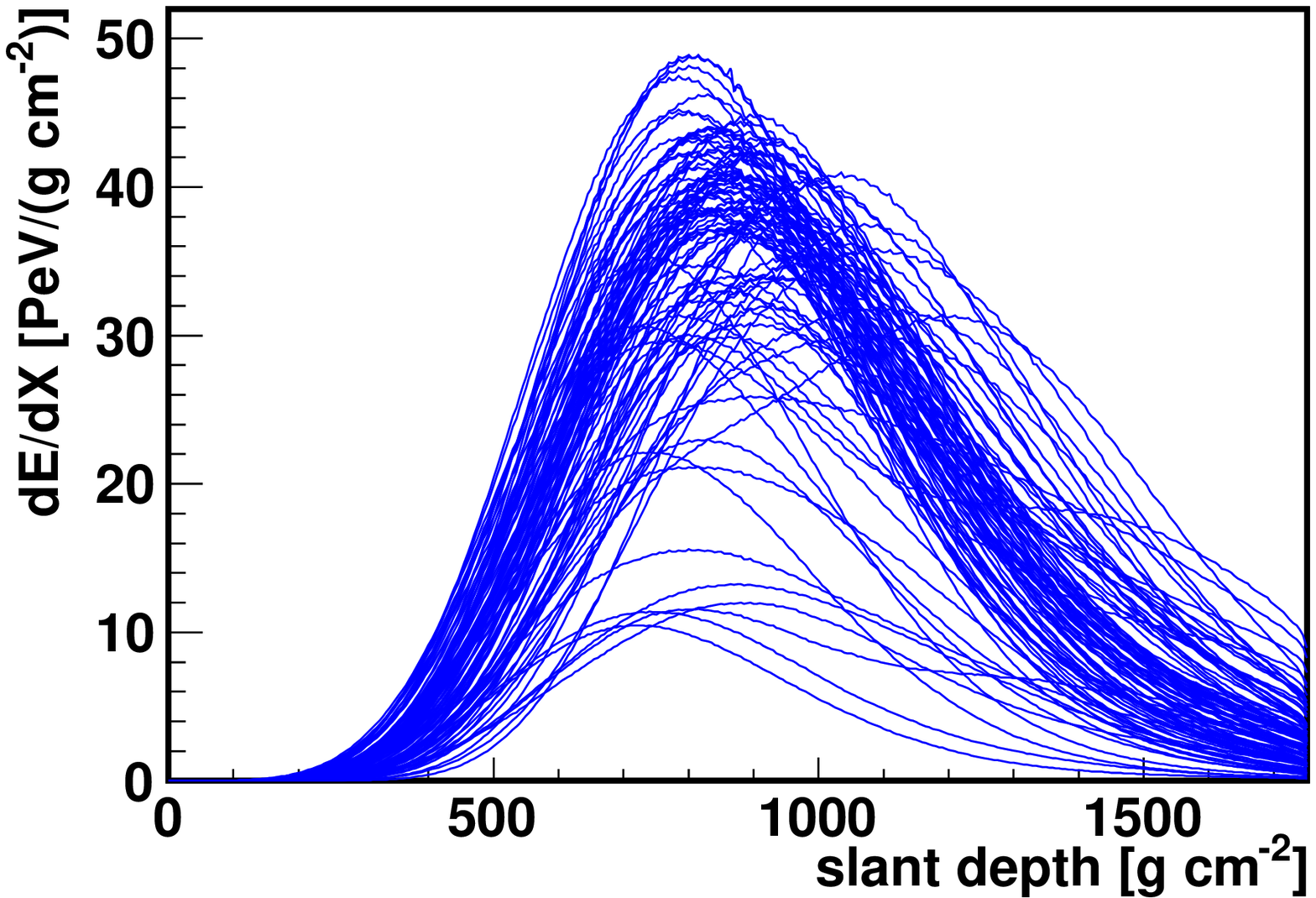}
    \label{fig:all_profilesB2}
  }	
  \subfigure[\textbf{highlighted profiles}]{
    \includegraphics[width=0.49\textwidth]{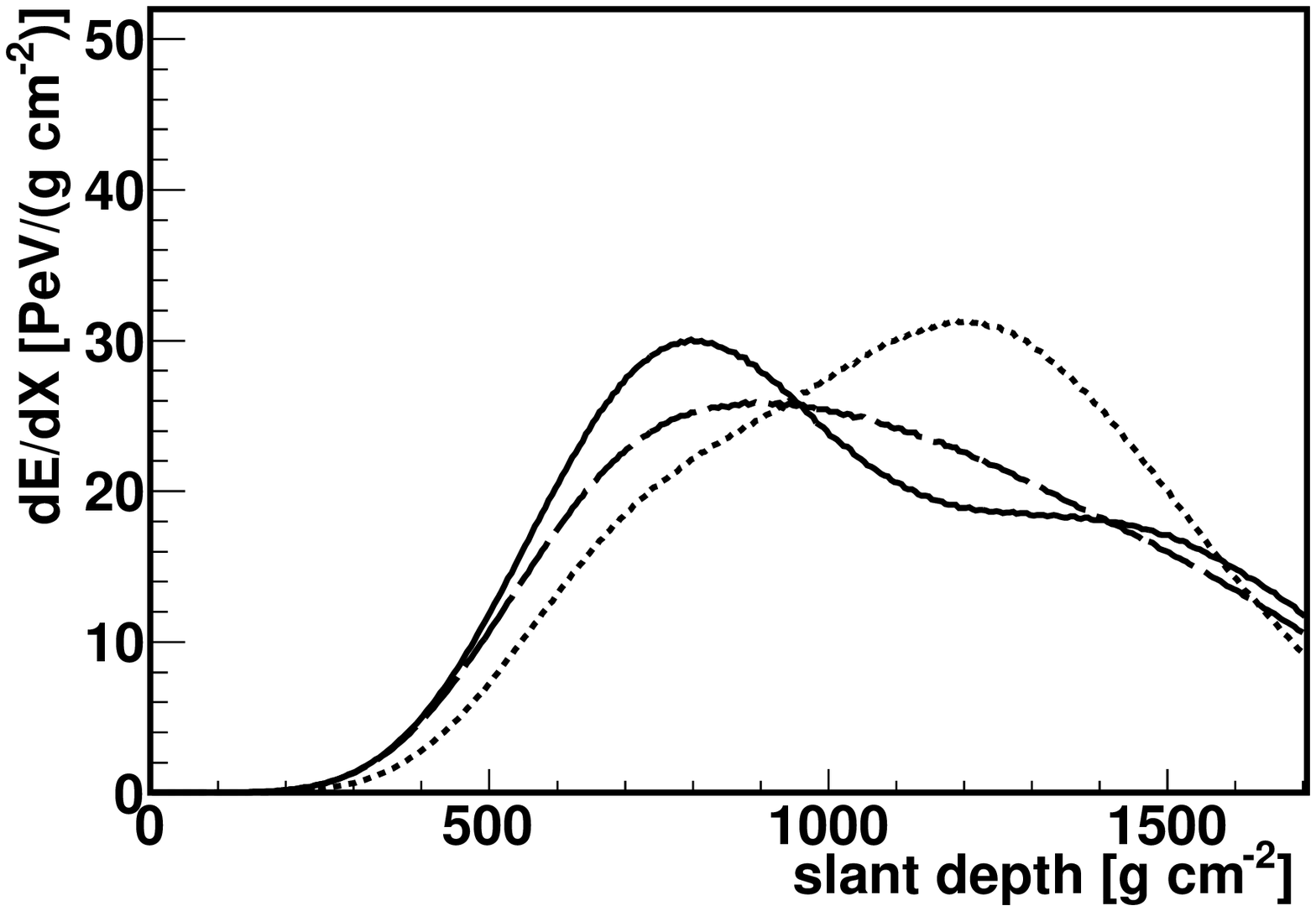}
    \label{fig:all_profiles_higlight}
  }
\caption{$10^{19.5}$ eV shower profiles for proton showers with no heavy quark production (a) and
with bottom quark production in two energy fraction bins ((b) and (c)). (d): Highlighted profiles from (c).}
\label{fig:all_profiles}
\end{figure}

To illustrate further how the energy of the leading heavy hadron
influences the development of the shower, we plot in
\figref{fig:decayB} and \figref{fig:decayC} 
the mean $10^{19.5}$ eV shower profiles for proton showers with charm
production and bottom production, respectively. We compared them 
to those of proton showers with no heavy quark production. We divide
the simulated events in two samples: one in which the heavy component
carries less than 50\% of the proton energy and its complementary set.
In the case of charm production, the average profile is only slightly different of that of proton showers.
However, in the case of bottom production, both samples are different
when compared to protons: they show a smaller number of particles at
maximum and on average they are 
deeper. We can see the differences more clearly if we inspect individual profiles.
In \figref{fig:all_profiles} we show the fluctuations in the shower profile due to the propagation of bottom hadrons. When 
the heavy component carries less than 50\% of the proton energy the effect is small, the shower profiles (\figref{fig:all_profilesB1}) resembling
those of proton showers with no heavy quark production (\figref{fig:all_profiles_proton}).
 For energy fractions above 50\% the effect is clearly noticeable (\figref{fig:all_profilesB2}). In \figref{fig:all_profiles_higlight}
 we highlight some showers from \figref{fig:all_profilesB2} whose profiles are specially anomalous. These showers 
show a slower development resulting either on a plateau or on a broader maximum. 

We can also consider extreme cases where the heavy hadron reaches
ground. This happens when numerous but very elastic interactions take place,
or if the hadron interacts only a few times with long distances
traveled between interactions. 
For EAS at $\theta$=60$^o$ the atmosphere has a slant depth of approximately 1760 g cm$^{-2}$. In \figref{fig:decayA} 
we plot the probability of decaying above 1700 g cm$^{-2}$ (equivalent to reaching ground) as a function of the production energy
 for different bottom hadrons (charmed hadrons have negligible probabilities of reaching ground, below 0.5\% at all energies, and are
not included in the plot). In those cases we have a standard
longitudinal profiles whose measured energy is smaller than that of
the primary particle, the rest of the energy being carried away by the heavy hadron
and not deposited in the atmosphere.

The discussion of whether the detection of heavy quarks in EAS is
feasible lies certainly beyond the scope of the present study. But we hope that some of the features
discussed in this section can be used in present or future cosmic ray
observatories to reveal that heavy hadrons are produced in the
atmosphere following the collisions of ultra-high energy cosmic rays. 

\section{Conclusions and prospects}
\label{sec:conclusions}

We discussed the modifications performed in CORSIKA to create and propagate heavy hadrons. We have written specific subroutines
to simulate their production during the first stage of an air shower development, and to treat their collisions with air nuclei.
These subroutines can be activated and deactivated through a set of
keywords in the CORSIKA control datacard. Our modifications have been
incorporated into the last official release of the CORSIKA package
(version 7.3500)~\cite{CorsikaVersion}.

The collisions of heavy hadrons with air turn out to be very elastic, with elasticity mean values above 50\% in all cases. If heavy hadrons are
produced with enough energy and they retain a high fraction of their
initial energy after a collision, they will interact again rather than
decaying. If a series of elastic interactions occur, the propagation
of the heavy hadrons will likely have observable effects on the shower development.

\section*{Acknowledgments}
We thank M. Masip and J.I. Illana for the theoretical input and discussions. We also thank 
D. Heck for his comments and reviews on the code modifications. The
assistance of G. Rubio and M.D. Serrano with shower simulations is warmly acknowledged. 
This work has been partially supported by MICINN of Spain (FPA2009-07187) and by Junta de Andaluc\'ia (FQM 330).
\appendix
\gdef\thesection{Appendix \Alph{section}}

\section{Particles considered and particle codes}
\label{sec:code:MonteCarlolist}
The bottom quark is not considered in CORSIKA because none of the hadronic interaction models
it implements produces bottom hadrons. 
If we want CORSIKA to propagate bottom hadrons, we first have to include them as eligible particles.
The last particle included is $\bar{\Sigma_c^*}^0$, with code 173. We
use the empty codes starting from 176 to include the new bottom hadrons. Bottom mesons and their antiparticles
are identified by codes from 176 to 183. $\Lambda$, $\Sigma$, $\Xi$ and $\Omega$ baryons and their
antiparticles have codes from 184 to 197. Only ground states of the particles have been introduced.
Details on the particle codes, masses and lifetimes, obtained from the
Particle Data Book \cite{PDG}, are shown in \tabref{tab:codes}.
\section{Input file}
\label{app:keywords}

CORSIKA reads a series of keywords to select the parameters of the simulations. These keywords have to be provided by the user as an input file. 
What follows is an example of a simple input file: in addition to the
standard keywords, we have underlined those keywords needed to use the
subroutines that control the physics of heavy hadron production and propagation:

\begin{lstlisting}
  RUNNR 1 number of run
  EVTNR 100400 no of first shower event
  SEED 100401 0 0 seed for hadronic part
  SEED 100402 0 0 seed for EGS4 part
  |\underline{COLLDR 1 3 }|
  |\underline{SIGMAQ 0 0 0 0}|
  |\underline{PROPAQ 1}|
  NSHOW 10 no of showers to simulate
  PRMPAR 14 primary particle code (proton)
  ERANGE 1.00E10 1.00E10 energy range of primary (GeV)
  THETAP 60. 60. range zenith angle (deg)
  PHIP -180. 180. range azimuth angle (deg)
  EXIT
\end{lstlisting}

\begin{itemize}
  
 \item \textbf{COLLDR} determines the type of the heavy quarks produced during the first interaction (first argument), and the production mechanism
 (second argument). The first argument accepts the values 1 for charm
 production, 2 for bottom production or 0 in case the first
 interaction is simulated by the chosen hadronic interaction model (be
 it SIBYLL, QGSJET, ...). The second one takes the values 1 for production via the Color
 Glass Condensate model, or 3, for production using the Intrinsic Quark model.
 \item \textbf{SIGMAQ} takes four arguments, the cross-sections (in
   mb) for interaction with protons of charmed mesons, charmed baryons, bottom mesons and bottom mesons, 
respectively. If the values are equal to 0 the parameterization shown in figure 3 is used.
 \item \textbf{PROPAQ} toggles the propagation of heavy hadrons with the new subroutines.
 If equal to 0, the propagation of heavy hadrons is performed by the
 high energy interaction model. If equal to 1, the propagation is dealt with using \textbf{HEPARIN}.
\end{itemize}

\section{Source code modifications}
\label{sec:code:code}

New functions have been written to perform specific parts of the simulation and some others have been modified inside 
 the source code to allow the propagation of the new particles. We list the files that have been modified,
those new files added, and overview the changes made to the code.\newline

The directory \textbf{corsika-6990/src/} contains the main source files needed to run CORSIKA. Inside the file \textbf{corsika.F}
we have made several modifications to already present subroutines:
\begin{itemize}
 \item \textbf{DATAC} reads the CORSIKA input file. This subroutine has been modified to accept the new keywords described
 in \ref{app:keywords}.
 \item the subroutine \textbf{NUCINT} selects the type of interaction process according to the particle energy.
 Now it includes a call to the new subroutine \textbf{COLLIDE}, to simulate
 the first interaction with production of heavy hadrons. The selection of interaction or
 decay routines for different particles types is extended to treat bottom hadrons.
 Both charmed and bottom hadrons interactions are treated in the new subroutine \textbf{HEPARIN}.
 \item \textbf{PAMAF} initializes the masses in GeV,
 the electric charge in electron charge units and  the mean life-times in s
 of the particles defined in CORSIKA. We modify it to hold the bottom hadrons defined in \ref{sec:code:MonteCarlolist} as well.
 \item \textbf{BOX2} determines the point of interaction or decay for any particle. It now uses the interaction 
 cross-sections of charmed particles with air shown in \figref{fig:crossint} to calculate their interaction lengths and whether they decay
 or interact. It has been extended to treat bottom hadrons as well.
 \item \textbf{PYTSTO} transports the particles resulting from PYTHIA to the CORSIKA stack. It is
 modified to accept bottom hadrons too.
\end{itemize}
We have also added new subroutines:
\begin{itemize}
 \item \textbf{HEPARIN} links with the PYTHIA routines
 that treat the interaction of heavy hadrons with air nuclei, instead
 of calling the high-energy model chosen during compilation.
 \item \textbf{NNY} samples the number of interacting nucleons in the collisions of heavy hadrons with air nuclei. 
 The sampled distributions are obtained using a modified version of NUCOGE \cite{NUCOGE}.
 \item \textbf{BTTMDC} is called to perform the decay of bottom hadrons.
\end{itemize}
Some of these modifications need the definition of new variables. These have been included in the header file
\textbf{corsika.h}.\newline

The file \textbf{qgsjet01c.f} simulates the physics of the model QGSJET01c. We have modified it to suppress the production of heavy quarks
during the first interaction. Thus, only \textbf{COLLIDE} (see below) produces them at that step of the shower. \newline

The directory \textbf{corsika-6990/pythia} contains all the PYTHIA routines called during the simulation of the shower. The 
source files of some of the new subroutines are here:
\begin{itemize}
\item the subroutine \textbf{COLLIDE}, in the file
  \textbf{collider.f}, produces the charmed or bottom hadrons at the
  first proton interaction.
We assume that the first interaction $p A \rightarrow H_{Q} H_{\bar{Q}} X$
  can be described as the superposition of the shower generated by the heavy hadrons
  ($H_{Q}$ and $H_{\bar{Q}}$) and the shower started by a proton of energy $E'_p = E_p-E_{H_Q}-E_{H_{\bar{Q}}}$.
  We use a proton as a primary and, once the depth of the first interaction ($X_0$) has
  been computed, we generate the pair $H_{Q}$ and $H_{\bar{Q}}$ at depth $X_0$, sampling the fractions of 
 the proton energy carried away ($x_1,x_2$) from the corresponding distributions. The energy
  of the proton is scaled to $E'_p$ and the proton shower starts at $X_0$. The particles are 
  transferred to the CORSIKA stack using the subroutine \textbf{PYTSTO}. The rest of the shower development follows the usual procedure.
  The type of particle produced (charm or bottom) and the production
 model (Color Glass Condensate or Intrinsic Quark) are chosen setting new keywords in the datacard (see \ref{app:keywords}).
 \item the subroutines \textbf{CHABADIF}, \textbf{CHABAPAR}, \textbf{CHAMEDIF}, \textbf{CHAMEPAR}, \textbf{BOBADIF},
 \textbf{BOBAPAR}, \textbf{BOMEDIF} and \textbf{BOMEPAR} (defined in the files with the same names and extension \textbf{.f}) 
 are called from \textbf{HEPARIN} to treat the diffractive  and partonic interactions of heavy hadrons. 
 The interactions are simulated according to the model described in section \ref{sec:physics}.
\end{itemize}

The processes included in the subroutines above need the modification
of two PYTHIA source files, pypdfu.f and pyspli.f.
\begin{table}[!t]
	\caption{CORSIKA particle codes extension. *: $\Sigma_b^0$, $\bar{\Sigma_b}^0$ are forced to decay whenever they are produced.}
	\begin{center}
		\begin{tabular*}{\textwidth}{@{\extracolsep{\fill}}| c | c | l | l || c | c | l | l |}
		\hline
		Particle & Particle & Particle & Particle & Particle & Particle & Particle & Particle\\
		code & name & mass & life-time & code & name & mass & life-time\\
		     &      &[GeV] &   [s]     &      &      & [GeV]& [s] \\
		\hline
		\hline
		176 & $B^0$ & 5.27958          & 1.519$\cdot10^{-12}$ & 187 & $\Xi_b^0$  &5.788              &  1.49$\cdot10^{-12}$\\
		177 & $B^+$ & 5.27925          & 1.641$\cdot10^{-12}$ & 188 & $\Xi_b^-$  &5.7911             &  1.56$\cdot10^{-12}$\\
		178 & $B^-$ & 5.27925          & 1.641$\cdot10^{-12}$ & 189 & $\Omega_b^-$  &6.071           &  1.1$\cdot10^{-12}$\\
		179 & $\bar{B}^0$ & 5.27958    & 1.519$\cdot10^{-12}$ & 190 & $\bar{\Lambda_{\textrm{b}}}^0$  &5.6194   &  1.425$\cdot10^{-12}$\\
		180 & $B_s^0$ & 5.36677        & 1.497$\cdot10^{-12}$ & 191 & $\bar{\Sigma_b}^+$  &5.8155    &  1.3$\cdot10^{-22}$\\
		181 & $\bar{B_s}^0$ & 5.36677  & 1.497$\cdot10^{-12}$ & 192 & $\Sigma_b^-$  &5.8113          &  1.68$\cdot10^{-23}$\\
		182 & $B_c^+$ & 6.277          & 0.453$\cdot10^{-12}$ & 193 & $\bar{\Xi_b}^0$  &5.788        &  1.49$\cdot10^{-12}$\\
		183 & $B_c^-$ & 6.277          & 0.453$\cdot10^{-12}$ & 194 & $\bar{\Xi_b}^+$  &5.7911       &  1.56$\cdot10^{-12}$\\
		184 & $\Lambda_{\textrm{b}}^0$ & 5.6194   & 1.425$\cdot10^{-12}$ & 195 & $\bar{\Omega_b}^+$  &6.071     &  1.1$\cdot10^{-12}$\\
		185 & $\Sigma_b^-$ & 5.8155    & 1.3$\cdot10^{-22}$   & 196 & $\Sigma_b^0$  & 5.8155         &  0 *\\ 
		186 & $\Sigma_b^+$  &5.8113    & 6.8$\cdot10^{-23}$   & 197 & $\bar{\Sigma_b}^0$  & 5.8155   &  0 *\\
		\hline                                                                       
		\end{tabular*}
		\label{tab:codes}
	\end{center}
\end{table}

\end{document}